\documentclass[twocolumn]{aastex63}
\usepackage{graphicx}
\usepackage{epstopdf}
\usepackage[utf8x]{inputenc}
\usepackage{xcolor}
\tabletypesize{\small}
\usepackage{wrapfig}
\usepackage{rotating}
\usepackage{natbib}
\usepackage{mathtools}
\usepackage[varg]{txfonts}
\usepackage{hyperref}
\usepackage{float}
\usepackage{mwe}
\defcitealias{ds2016}{DS2016}	

\begin{document}

\title{Lack of influence of the environment in the earliest stages of massive galaxy formation}

\author[0000-0002-8053-8040]{Marianna Annunziatella}
\affil{Centro de Astrobiolog\'ia, (CAB, CSIC-INTA), Carretera de Ajalvir km 4, E-28850 Torrejón de Ardoz, Madrid, Spain}
\affil{INAF-Osservatorio Astronomico di Capodimonte, Via Moiariello 16, 80131 Napoli, Italy}

\author[0000-0003-4528-5639]{Pablo G. P\'erez-Gonz\'alez}
\affil{Centro de Astrobiolog\'ia, (CAB, CSIC-INTA), Carretera de Ajalvir km 4, E-28850 Torrejón de Ardoz, Madrid, Spain}

\author[0000-0002-8365-5525]{\'Angela Garc\'ia Argum\'anez}
\affil{Departamento de F\'{\i}sica de la Tierra y Astrof\'{\i}sica, Faultad de CC. F\'{\i}sicas, Universidad Complutense de Madrid, E-28040, Madrid, Spain}

\author[0000-0001-6813-875X]{Guillermo Barro}
\affil{University of the Pacific, Stockton, CA 90340, USA}

\author[0000-0002-4140-0428]{Bel\'en Alcalde Pampliega}
\affil{European Southern Observatory,Alonso de Córdova, Santiago de Chile, Chile}

\author[0000-0001-6820-0015]{Luca Costantin}
\affil{Centro de Astrobiolog\'ia, (CAB, CSIC-INTA), Carretera de Ajalvir km 4, E-28850 Torrejón de Ardoz, Madrid, Spain}
\affil{INAF - Osservatorio Astronomico di Brera, Via Brera 28, 20121, Milano, Italy}
\author[0000-0002-6610-2048]{Anton M. Koekemoer}
\affil{Space Telescope Science Institute, 3700 San Martin Dr., 
Baltimore, MD 21218, USA}

\author[0000-0001-8115-5845]{Rosa M. M\'erida}
\affil{Centro de Astrobiolog\'ia, (CAB, CSIC-INTA), Carretera de Ajalvir km 4, E-28850 Torrejón de Ardoz, Madrid, Spain}

\keywords{galaxies:general, galaxies: formation, galaxies: evolution, galaxies: high-redshift, galaxies: interactions}

\begin{abstract}
We investigate how the environment affects the assembly history of massive galaxies. For that purpose, we make use of SHARDS and HST spectro-photometric data, whose depth, spectral resolution, and wavelength coverage allow to perform a detailed analysis of the stellar emission as well as obtaining unprecedentedly accurate photometric redshifts. This expedites a sufficiently accurate estimate of the local environment and a robust derivation of the star formation histories of a complete sample of 332 massive galaxies ($\mathrm{>10^{10}M_{\odot}}$) at redshift $1\leq z \leq 1.5$ in the GOODS-N field. We find that massive galaxies in this redshift range avoid the lowest density environments. 
Moreover, we observed that the oldest galaxies in our sample with  with mass-weighted formation redshift $\mathrm{\overline{z}_{M-w} \geq 2.5}$, avoid the highest density regions, preferring intermediate environments. Younger galaxies, including those with active star formation, tend to live in denser environments ($\Sigma = \mathrm{5.0_{1.1}^{24.8}\times 10^{10}M_{\odot}Mpc^{-2}}$).  This behavior could be expected if those massive galaxies starting their formation first would merge with neighbors and sweep their environment earlier. On the other hand, galaxies formed more recently ($\overline{z}_{M-w} < 2.5$) are accreted into large scale  structures at later times and we are observing them before sweeping their environment or, alternatively, they are less likely to affect their environment. However, given that both number and mass surface densities of neighbor galaxies is relatively low for the oldest galaxies, our results reveal a very weak correlation between environment and the first formation stages of the earliest massive galaxies.

\end{abstract}
\section{Introduction}\label{s:intro}

Our understanding of how galaxies form and evolve is still challenged by numerous observations. According to the standard $\Lambda CDM$ paradigm \citep{white1991}, the dominant structures in the universe are dark matter halos which grow out of an initial field of density perturbations. Simulations and analytical models show that this process proceeds primarily in a hierarchical, bottom-up manner, with low-mass halos forming early and subsequently growing via continued accretion and merging to form the most massive halos at later times \citep{kauffmann1993, baugh1998}. In contrast to the predicted hierarchical growth of the dark matter halos, different observational studies have shown that the stellar baryonic component of the halos (i.e., the galaxies) grow to some extent in an anti-hierarchical, top-down manner. Remarkably, it appears that the most massive galaxies in the local universe assembled the majority of their stellar mass rapidly and are already present in significant numbers at early times \citep[e.g.,][]{perez-gonzalez2008, marchesini2010, marchesini2014, alcalde2019, forrest2020}. Even though many massive galaxies at $z>1$ show significant recent star formation,  there is a numerous subpopulation of these objects with already old stellar populations at those early epochs.

While, thanks to the advent of wide and deep NIR surveys, massive ($\mathrm{> 10^{10}M_{\odot}}$) galaxies have been discovered now in larger numbers across different redshifts, a clear understanding of the formation mechanisms of this population is still missing. In particular, an accurate description of the star formation histories (SFHs) of massive galaxies is crucial to understand how the universe has evolved. 

Specially suited to get robust SFHs, spectroscopic observations are, however, time consuming, available for limited samples of galaxies and biased towards the brightest objects \citep[e.g., ][]{belli2021,forrest2020, schreiber2018}. On the other hand, the spectral resolution of broad-band photometric surveys is not enough to perform a detailed analysis of the stellar population properties due to the degeneracies in the analysis of their spectral energy distributions (SEDs) using stellar population models.

In \citealt{ds2016} (DS2016, hereafter), the authors took advantage of the spectrophotometric resolution and depth of the Survey for High-z Absorption Red and Dead Sources survey \citep[SHARDS; ][]{perez-gonzalez2013}, together with grism observations from the CANDELS and AGHAST/3D-HST surveys \citep{grogin2011, koekemoer2011, brammer2012}, to obtain robust estimations of the SFH of a sample of massive quiescent galaxies at $1<z<1.5$. 

\citetalias{ds2016} found that  the population of massive quiescent galaxies at $1.0<z<1.5$ is dominated by 1~Gyr old galaxies. A small fraction (15\%) of the whole populations corresponded to old galaxies (ages older than 2~Gyr), which were identified with the descendants of $z\sim2$ massive quiescent galaxies \citep[also known as {\it red nuggets}; ][]{vandokkum2006, barro2013, vandokkum2015}. These results were also found by other teams using mainly spectroscopic data (among others, see \citealt{choi2014,schreiber2018,belli2019,morishita2019}).
In this paper, we extend the SFH analysis in \citetalias{ds2016} to all massive galaxies at $1.0<z<1.5$ identified in the SHARDS survey in the GOODS-N field. Furthermore, we take advantage of the unprecedented accuracy of the photometric redshift obtained using SHARDS and grism data (mean $\mathrm{\Delta z/(1+z) = 0.0023 }$, \citealt{barro2019}), to investigate the interplay between the environment and the SFHs of massive galaxies at $z>1$.

In the nearby Universe, the interplay between galaxy properties and environment is well studied \citep[e.g.][]{dressler1980, blanton2005, pa2020}. The general picture that emerges at low redshift is that in the more dense environment, i.e. cluster of galaxies, galaxies are predominantly redder, have an early type morphology, form less stars, and host older stellar populations \citep[for a review, see][]{silk&manon2012}. Not only the fraction of quiescent galaxies is higher in clusters, but the level of star formation activity of star-forming galaxies is lower with respect to the field \citep{Elbaz_2007,vulcani2010}. 
Beyond the nearby Universe, the analysis of environment has been, so far, severely affected by the scarcity of spectroscopic redshifts and the too large uncertainties on photometric redshifts. However, in the last decade some studies have emerged across a range of redshifts.
At z$\sim$1, \cite{marcillac2008} found that luminous infrared galaxies tend to avoid very underdense environments and favour environments 1.2 times denser than the mean local environments of galaxies at the same redshift. At the same redshift, \cite{popesso2011} confirmed an anti-correlation between the level of star-formation and the environment found at lower redshift. At $2<z<5$, \cite{lemaux2022} observed a definite, nearly monotonic increase in the average star-formation rate with increasing galaxy overdensity, due to the increase in high-density environments of galaxies  that are more massive in their stellar content and are forming stars at a higher rate than their less massive counterparts. The trend is still present, but weakened, when stellar mass effects are accounted for.
At z$\sim$ 2, \cite{strazzullo2013} found environmental signatures on galaxy populations in the most central region of clusters, which presented a clear increase in the passive fraction of massive galaxies. \cite{cucciati2010}, found that 
the fraction of red  galaxies depends on the environment at least up to z $\sim$ 1, with red galaxies residing mainly in high densities, however they also found that this trend becomes weaker for increasing redshifts.
At $z>1$, \cite{muzzin2012} investigated the relation between specific star-formation rates, mass, $D_n$4000 and environment in cluster and field star-forming and quiescent galaxies. They found that at fixed stellar mass, both sSFR and  $D_n$4000 are independent of environment.
\citealt{raichoor2011,rettura2011} find that at $z \sim 1.3$ early type galaxies of similar mass in cluster and in the field have similar ages.  More recently, \cite{webb2020} found a 0.3 Gyr difference in the mass-weighted ages of cluster and field quiescent galaxies at $1.0 < z < 1.5$ only for galaxies with $10.5 < log(M/M_{\odot}) < 11.3$. At slightly lower redshift ($0.6<z<1.0$), \cite{sobral2022} found that quiescent galaxies in overdensities have higher $D_n$4000 and lower H$\delta$, indicating older stellar populations with respect to galaxies in lower density environments. For star-forming galaxies, there is no difference of $D_n$4000 and H$\delta$ with environment. On the other hand, at $0.5 \leq z \leq 1$, \cite{moresco2010}, find that the age difference for early type galaxies in different environment is very small ($<$ 0.2 Gyr).
\\
A complementary way to understand how galaxies are affected by the environment in which they reside is to look at the two-point correlation function (2PCF) \citep{davisp1983} of different types of galaxies. This method is widely used to compare observational results to mock galaxy simulations. While it would be preferable to measure the 2PCF using spectroscopic redshifts, many works at high redshift use photometric redshifts, with the same limitations as in environmental studies. 
\cite{foucaud2010} derived the 2PCF for massive galaxies ($>10^{10}M_{\odot}$) at different redshifts up to $z=2$, and found that the most clustered systems are galaxies at the highest stellar masses with $10^{11}M_{\odot} < M_{\star} < 10^{12}M_{\odot}$ at $z\sim $2. At $z\sim 1$, \cite{mostek2013} found a high clustering amplitude for red galaxies with high stellar masses and for blue galaxies with high star formation rates. At $z<0.5$, \citet{sureshkumar2022} observed that galaxies brighter in the mid-infrared exhibit stronger clustering than their fainter counterparts.\\
In this work, thanks to the deep spectro-photometric data in our sample we are able to characterize the environment of galaxies at $1.0 \leq z \leq 1.5$ in the GOODS-N field with unprecedented accuracy and to study how it may affect the galaxies' star formation histories. 
The paper is organized as follows:
in Section \ref{s:sample_sel}, we briefly describe the data and present the sample selection. In Section \ref{ss:sed_model}, we outline the SED fitting method, referring the reader to more details in \citetalias{ds2016}. In Section \ref{s:properties}, we derive the mean properties of the galaxies in our sample, while in Section \ref{s: env} we describe different estimations of the environment and show how the galaxy properties depend on it. Finally, in Section \ref{s:disc}, we draw our conclusions.

Throughout the paper we assume a flat cosmology with $\mathrm{\Omega_M\, =\, 0.3,\, \Omega_{\Lambda}\, =\, 0.7}$, and a Hubble constant $\mathrm{H_0\, =\, 70\, km s^{-1} Mpc^{-1}}$, and use AB magnitudes \citep{okegunn1987}. All stellar mass and SFR estimations refer to a \citet{chabrier2003} IMF.

\section{Sample Selection and characterization}\label{s:sample_sel}

Our sample was selected from the spectro-photometric catalog described in \cite{barro2019} of galaxies in the GOODS-N region. The photometric catalog was constructed using the CANDELS WFC3/F160W detection and comprises an extensive ancillary dataset spanning from the ultraviolet (UV) to far-infrared (FIR) wavelengths. 

Particularly relevant to this paper is the availability of medium band photometry from the SHARDS survey \citep{perez-gonzalez2013}, which imaged 110 $\mathrm{arcmin^2}$ of the GOODS-N region with 25 contiguous medium-band filters in the wavelength range between 500 and 950 nm, leading to a spectral resolution R$\sim$50. The SHARDS data reach a depth of H$=$27 AB at 3$\sigma$ with subarcsecond seeing in each single band.

Also essential for this work, our work includes spectroscopic data in the NIR consisting in WFC3 G102 and G141 grism observations from the AGHAST survey \citep{weiner2009}. Grism observations grant a continuous coverage between $\mathrm{\lambda\,=\,0.8\, \mu m}$ to $\mathrm{\lambda\,=\,1.7\, \mu m}$ and with a resolution $\mathrm{R\,=\,130}$. In summary, the combination of SHARDS and AGHAST data provides spectrophotometric data from 0.5 to 1.7~$\mu$m with spectral resolution $R\gtrsim50$ or better, which expedites both an accurate and unprecedented determination of photometric redshifts as well as stellar population properties.

The catalog from \cite{barro2019} also gathers information such as photometric and spectroscopic, if available, redshifts, and the total star-formation rate (SFR), obtained by combining UV and IR indicators. Ancillary data include also IRAC and NIR K-band photometry. All this information has been extracted from the Rainbow database\footnote{\url{http://arcoirix.cab.inta-csic.es/Rainbow_navigator_public/}.} \citep{perez-gonzalez2008, barro2011}.

The use of the SHARDS medium-band data makes the accuracy of the photometric redshift presented in this catalog unprecedentedly high, varying from $\mathrm{\sigma_{nmad}\, = 0.0028}$ for galaxies with broad-band and SHARDS photometry to $\mathrm{\sigma_{nmad}\, = 0.0023}$ for galaxies that have grism spectra \citep{barro2019}, up to redshift $z=3$ and $I\lesssim25$~mag (for which spectroscopic data is available). If we only consider galaxies at $1.0<z<1.5$, our redshift range of interest, these numbers are even better: median $\mathrm{\sigma_{nmad}\, = 0.0021}$.

Using the SHARDS/CANDELS dataset from \citet{barro2019}, we selected a sample of massive galaxies, i.e., with $\mathrm{log(M/M_{\odot}) \geq 10}$, at $\mathrm{1\leq z\leq 1.5}$. We visually inspected all the galaxies and removed a few problematic sources (3\% of the total sample), like misclassified stars or objects too close to bright sources that could have contaminated photometry. The final sample comprises 332 galaxies over the entire redshift range.

\begin{figure*}[h]
\begin{tabular}{cc}
\multicolumn{2}{c}{
 \includegraphics[width=0.6\textwidth]{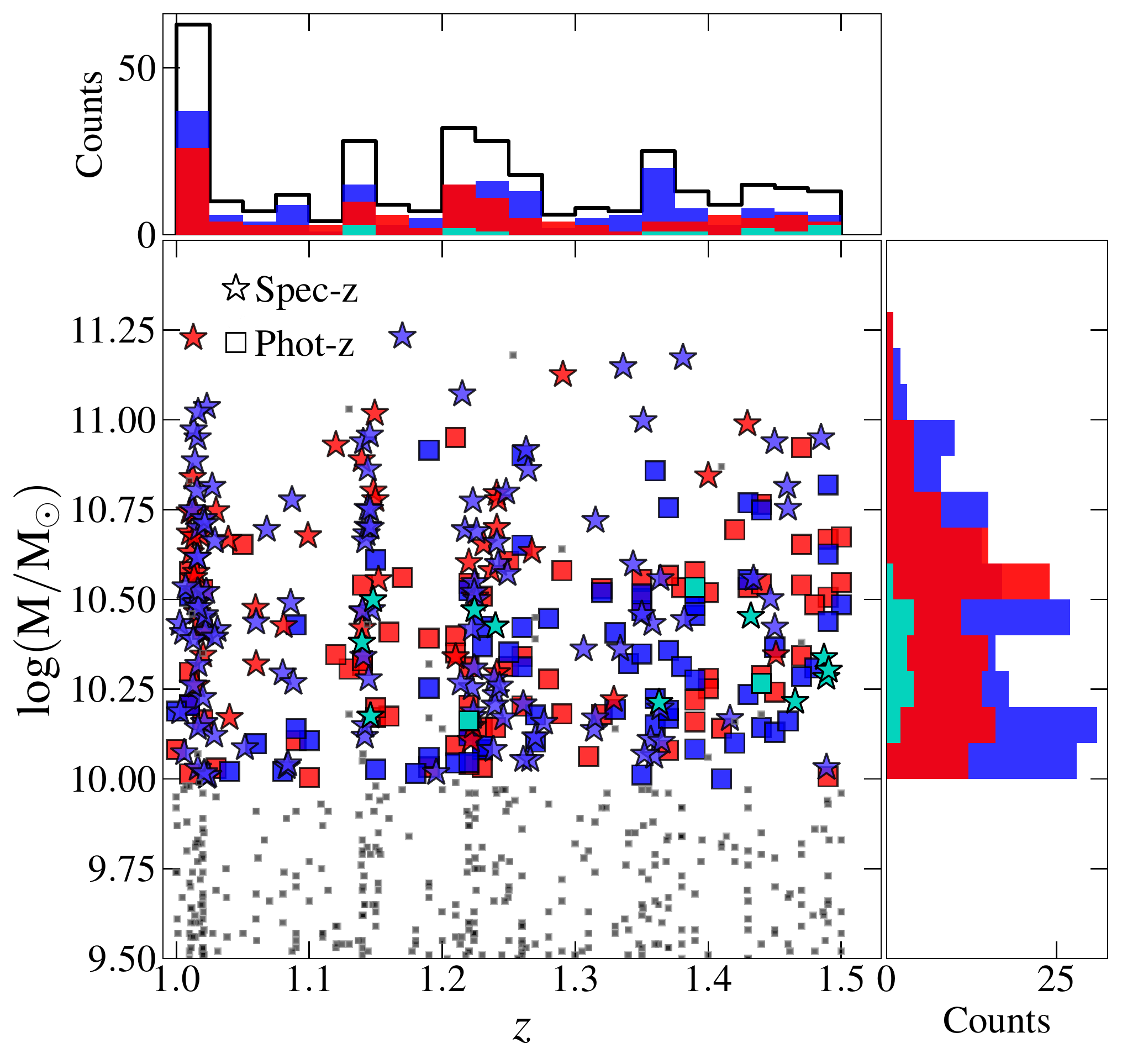} } \\
\multicolumn{2}{c}{(a)}\\
  \hspace*{-1cm}%
  \includegraphics[width=0.523\textwidth, trim= 0 5 5 -20,clip]{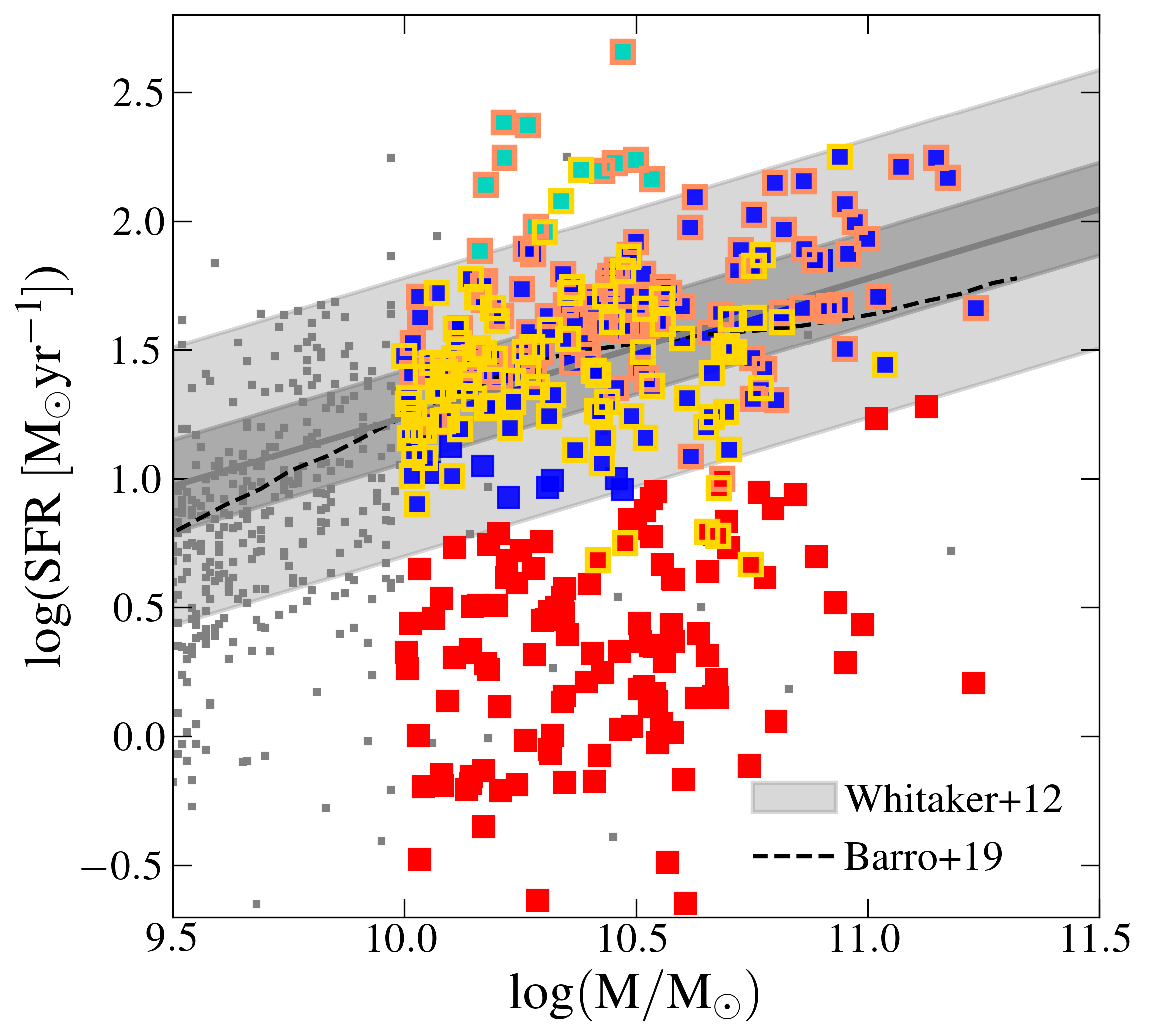} & \includegraphics[width=0.5\textwidth, trim= 0 5 5 -5,clip]{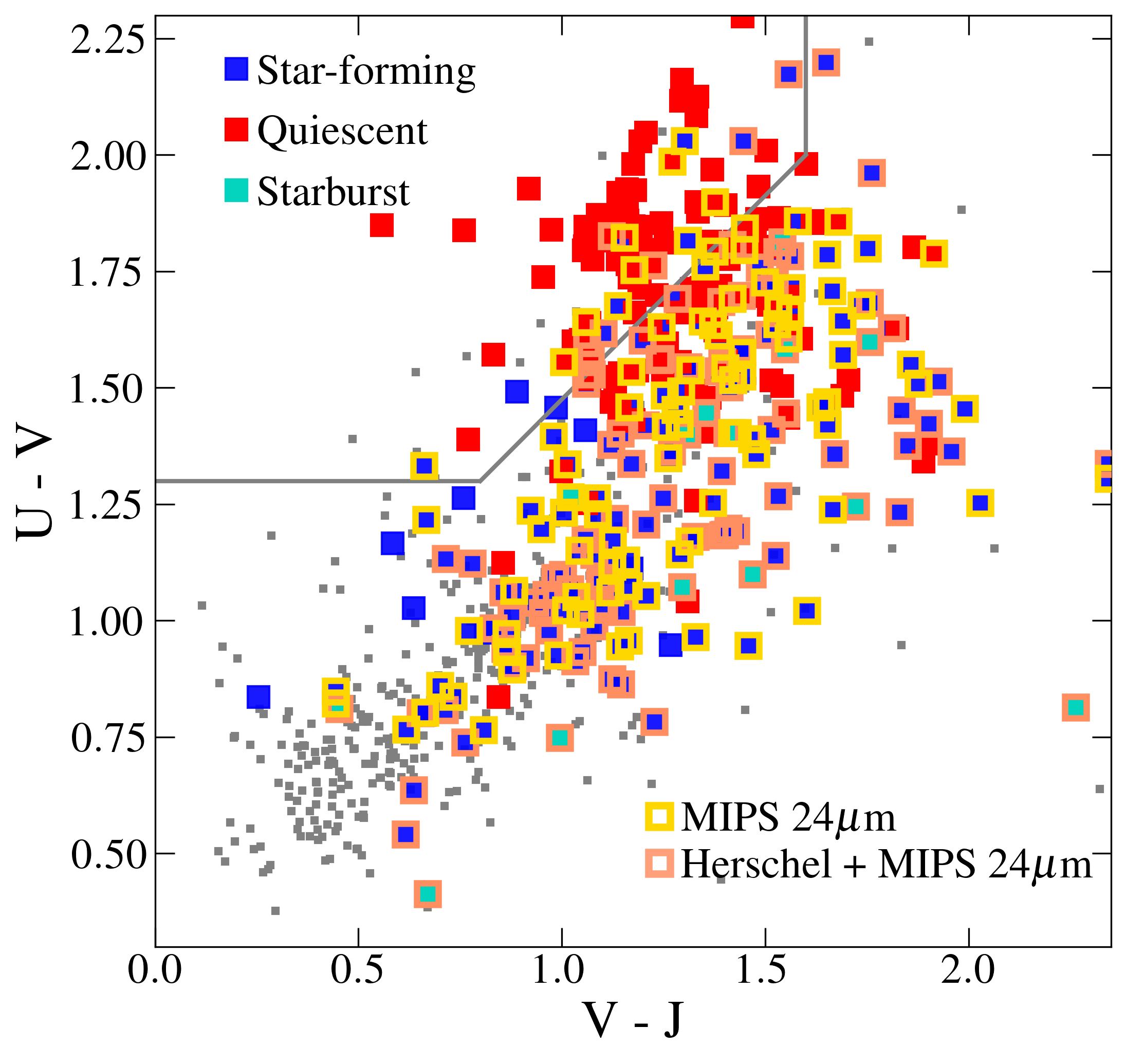} \\
  (b) & (c) 
\end{tabular}

\caption{Panel (a) Mass versus redshift distribution of the entire SHARDS/CANDELS catalog (gray points) and the selected sample (color-coded according to the distance to the coeval SFR main sequence from \citep{whitaker2012}. Galaxies plotted as a star have a spectroscopic redshift (54\% of the sample), squares refer to galaxies with a photometric redshift. Upper panel: redshift distribution of all massive galaxies (in black), massive star-forming galaxies (blue), starburst galaxies (cyan), and massive quiescent galaxies (red). 
Panel (b): Star formation rate vs stellar mass relation for all galaxies in the CANDELS GOODS-N catalog (gray points) and our sample (colored points). The MS from \citet{whitaker2012} is  plotted, with its 1$\mathrm{\sigma}$ (dark grey area) and 3$\mathrm{\sigma}$ (light grey area)  scatter. We also show the running median of the MS from \citet{barro2019}. Galaxies detected in MIPS 24 $\mu$m are marked in orange, while galaxies detected also in HERSCHEL 100-500 $\mu$m are marked in red. Star-forming galaxies are plotted in purple, quiescent galaxies in red and starburst galaxies in blue. Panel (c): $UVJ$ diagram of the same samples shown in previous plots. Colors are the same as in Panel (b).}
\label{f:mass_z}
\end{figure*}

In Fig.~\ref{f:mass_z}(a), we show the mass versus redshift plot for the entire SHARDS + CANDELS dataset, down to the mass completeness limit of $\mathrm{10^{9.5}\, M_{\odot}}$ at $z$=1.5 \citep[black squares,][]{barro2019}. Our selected sample is color-coded according to the distance from the coeval SFR main sequence \citep{whitaker2012}, which is shown in panel b. Throughout the paper, we used perceptually uniform color-maps \citep{kovesi2015}. In Fig.~\ref{f:mass_z}(b), we show the SFR vs mass diagram for the entire sample (black squares) and for the selected sample (blue points). We overplot the star-forming main sequence (MS) derived by \citep{whitaker2012} and \citep{barro2019} at this redshift and their quoted 1$\mathrm{\sigma}$  scatter. Based on this plot and herafter, galaxies that are more than 3$\mathrm{\sigma}$ below the main sequence are classified as quiescent. The rest of the galaxies are classified as star-forming. Fig.~\ref{f:mass_z}(c) shows the $\mathrm{U-V}$ vs $\mathrm{V-J}$ color diagram ($UVJ$) of the same samples.

Based on Fig.~\ref{f:mass_z}, we infer that 27\%  and 12\% of our sample of 332 massive galaxies are located in 2 prominent redshift peaks at $z=1.02$ and $z=1.13$, respectively. Another wider concentration of galaxies is found at $z=1.23$. Around 38\% of the sample lies below 3$\sigma$ from the MS and is classified as quiescent galaxies. Only 24\% of the galaxies would be classified as quiescent according to the $UVJ$ diagram. 
Just 4\% of our galaxies qualify as strong starbursts (at least $3-\sigma$ above the MS). The majority (85\%) of our galaxies present red colors, $U-V>1.0$ mag and $V-J>1.25$ mag. 

\subsection{SED modeling}\label{ss:sed_model}

To characterize in detail the star-formation histories (SFHs) of massive galaxies at $\mathrm{1.0\leq z \leq1.5}$, we fit the observed photometric and grism data to stellar population synthesis (SPS) models using the fitting code {\it{synthesizer}} described in \cite{perez-gonzalez2003,perez-gonzalez2008}. We chose to model the galaxy SFHs with a single burst of star formation characterized by a delayed exponentially declining law:
\begin{equation}
   \mathrm{SFR(t)\,\propto\, t/\tau^2\, \times e^{-t/\tau}.}
\end{equation}
This parametrization was chosen in order to avoid the non-physical infinite derivative at $t\,=\,0$ obtained in the simple exponential decay law. We compare the SEDs of our galaxies with the BC03 stellar population library \citep{bc03}, assuming a \cite{chabrier2003} initial mass function (IMF) and the \cite{calzetti2000} attenuation law. The values of the $V$-band attenuation vary between 0 and 4~mag. 
We allow four discrete metallicity values $Z/Z_{\odot} \, =\, [0.2,0.4,1,2.5]$, two sub-solar, solar and super-solar, to account for the fact that massive galaxies tend to be more metal rich \citep{gallazzi2005, gallazzi2014, saracco2020}. Furthermore, we consider a minimum value for the star formation timescale $\tau=100$~Myr and impose that the SFH starting time ($t_0$) for a given galaxy does not exceed the age of the Universe at its redshift. A summary of the free parameters and their allowed ranges in the SED-fitting procedure is reported in Table~\ref{t:parameters}.

\begin{deluxetable}{cccc}
	\tablewidth{0pt}
	\tablecaption{Free parameters and their allowed ranges used in the SED-fitting procedure.
		\label{t:parameters}}
	\tablehead{\colhead{Parameter} & \colhead{Range}  & \colhead{Units} }
	\startdata 
	\hline
	SFH starting time ($t_0$) & 0.001 – 5.75 & Gyr \\
	Time-scale ($\tau$) & 100 - 100000 & Myr \\
	Dust attenuation ($\mathrm{A_V}$) & 0 - 4 & mag \\
	Metallicity (Z) & 0.2, 0.4, 1.0, 1.5 & $\mathrm{Z_{\odot}}$\\
	\hline
	\enddata
\end{deluxetable}

The \textit{synthesizer} code performs a $\chi^2$ minimization to find the best-fitting model.
To deal with possible degeneracies in the solutions, we use the Monte-Carlo approach described in \citetalias{ds2016}, and determine full probability distributions for each fitted parameter. We constructed 1000 modified SEDs by allowing the spectro-photometric data points to randomly vary following a Gaussian distribution, with a width given by the photometric errors. We perform the SED fitting of all the 1000 realizations and obtain the corresponding set of parameters for every galaxy in our sample. Then, we look for clusters of solutions in the $\tau-t_0$ parameter space. Clusters of solutions which provide similar results are grouped as a single solution identified by a median value and a scatter in the multi-dimensional $\tau-t_0-A_V-Z$ space. We then assign a  statistical significance to each cluster according to the fraction of solutions belonging to it.
To break the degeneracies identified in the $\tau-t_0$ parameter space, we take advantage of the spectrophotometric data from SHARDS and the grism observations. As explained in detail in \citetalias{ds2016}, we derive the $\mathrm{mg_{UV}}$ and $D4000$ indices, which are well correlated to stellar population ages \citep{kauffmann2003, daddi2005}. We measure these indices from the spectrophotometric data alone, hence they are independent of the SED fitting. We use the measurements of these indices to eliminate solutions which are incompatible with the spectrophotometric data. This procedure can be interpreted as an increased weight given in our SED-fitting method to some useful absorption indices which are well correlated to the most interesting physical property that we want to extract: the age.  
In addition, we use an energy balance argument in the SED-fitting to discard impossible solutions. For each best-fitting solution, we derive the expected flux at an observed wavelength of 24 $\mu$m ($\mathrm{F_{pr}}$(24)). To do that, first we calculate the luminosity absorbed by dust in the UV/optical according to a given best-fitting model. This absorbed energy has to be re-emitted in the IR. We assume that the stellar emission absorbed by dust must be equal to the IR luminosity integrated from 8 to 1000 $\mu$m ($\mathrm{L_{IR}}$). Then, we use the relation in \cite{rujopakarn2013} to convert $\mathrm{L_{IR}}$ to the 24 $\mu$m observed flux at the appropriate redshift. The detection limit in 24 $\mu$m for our data is $\sim$50~$\mu$Jy at 90\% completeness (\citealt{pg2005}, \citealt{barro2019}). For galaxies undetected in the IR, we can discard solutions with predicted 24 $\mu$m fluxes larger than the detection limit. An example of the full fitting and solution purging method is shown in Appendix \ref{a:sed}.

\section{Sample properties}\label{s:properties}

In this section, we analyze the statistical properties of the stellar populations in massive galaxies at $\mathrm{1.0\leq z \leq1.5}$. 
\subsection{Mass weighted ages}\label{ss:tM}

\begin{figure*}[ht]
    \centering
    \includegraphics[scale=0.54]{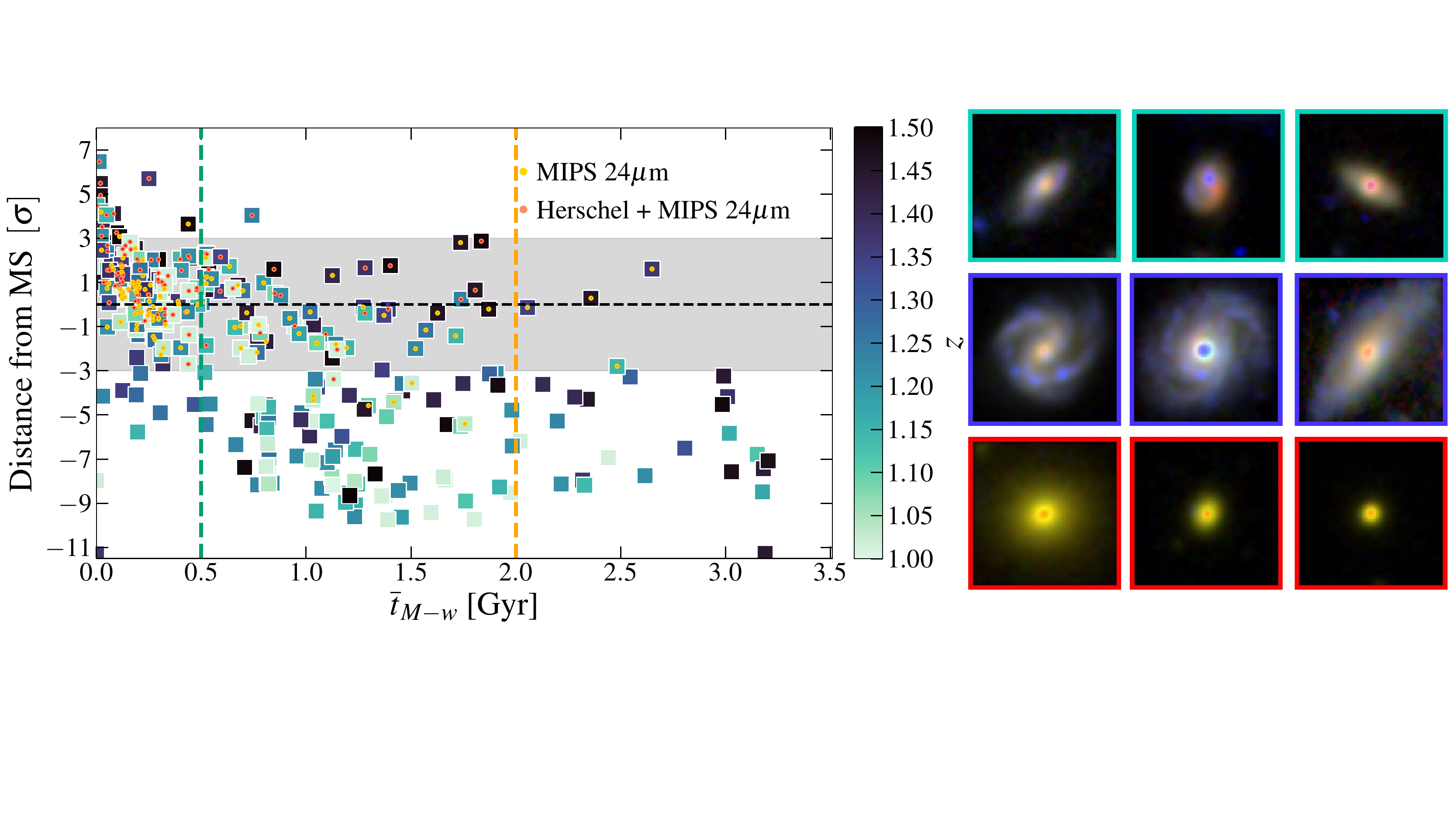}
    \caption{Left : Distance from the SFR-stellar mass main sequence at $1.0<z<1.5$ as a function of the galaxy mass-weighted age. Galaxies detected in MIPS 24$\mu$m are marked in orange, while galaxies detected in HERSCHEL 100-500 $\mu$m are marked in red. The green and orange dashed lines  mark the separation between  young and mid-age and mid-age and old galaxies, respectively. The horizontal dashed line mark the position of the MS from \citealp{whitaker2012}. Data points are also color coded by redshift. Right: 5\arcsec $\times$5\arcsec RGB postage stamps obtained from HST images (F814w, F125w and F160w) for three starburst (cyan), star-forming (blue), and three quiescent (red) galaxies. The bluer HST images have been convolved with a gaussian kernel with a FWHM matching the resolution of F160W. Then to each band a factor equal to the ratio of the integral of the filter profile to a top-hat profile with the same width was applied. The starburst galaxies in this sample are all very dusty, detected both in MIPS 24$\mu$m and in HERSCHEL 100-500 $\mu$m.}
    \label{f:tm_ms}
\end{figure*}

For all galaxies in our sample, we derived the mass-weighted ages as:
\begin{equation}
    \overline{t}_{M-w}=t_0-\frac{\int_{0}^{t_0} SFR(t)\times t dt}{\int_{0}^{t_0} SFR(t)dt}=t_0-\frac{\int_{0}^{t_0} SFR(t)\times t dt}{M},
\end{equation}
where $t_0$ is the SFH starting time, $M$ is the best-fitting mass, and the {\it {t}} is the time since the start of the star formation in the galaxy. While $t_0$ corresponds to the beginning of the star formation period that dominates the SFH of the galaxy, the mass-weighted age is a better approximation to the average age of the stellar populations in a given galaxy and takes into account the tau-age degeneracy.
Fig.~\ref{f:tm_ms} shows the distance of the galaxies in our sample from the coeval star-formation main sequence \citep[MS,][]{whitaker2012, barro2019} as a function of the mass-weighted age. Mid-infrared (MIR) and FIR emitters are marked in the plot. The SFR for each galaxy is a combination of mid and FIR and UV indicators as presented in \citet{barro2019}, and it is not a direct product of the SED fitting. The plot shows the consistency between our SPS modelling and SFRs determined with observables. 

Based on this plot, we decided to divide our sample in three classes according to their mass-weighted age. In Fig.~\ref{f:tm_ms}, we show the RGB color images of three galaxies for each class, defined as:

\begin{itemize}
    \item galaxies with $\overline{t}_{M-w}<$ 0.5 Gyr (young galaxies, hereafter),
    \item galaxies with 0.5 $\leq\overline{t}_{M-w}<$ 2 Gyr (mid-age galaxies, hereafter),
    \item galaxies with $\overline{t}_{M-w}\geq$ 2 Gyr  (old galaxies, hereafter).
\end{itemize}

\begin{figure*}[ht]
\begin{tabular}{cc}
  \hspace*{-1cm}%
    \includegraphics[width=0.52\linewidth, trim= 0 5 5 -5,clip]{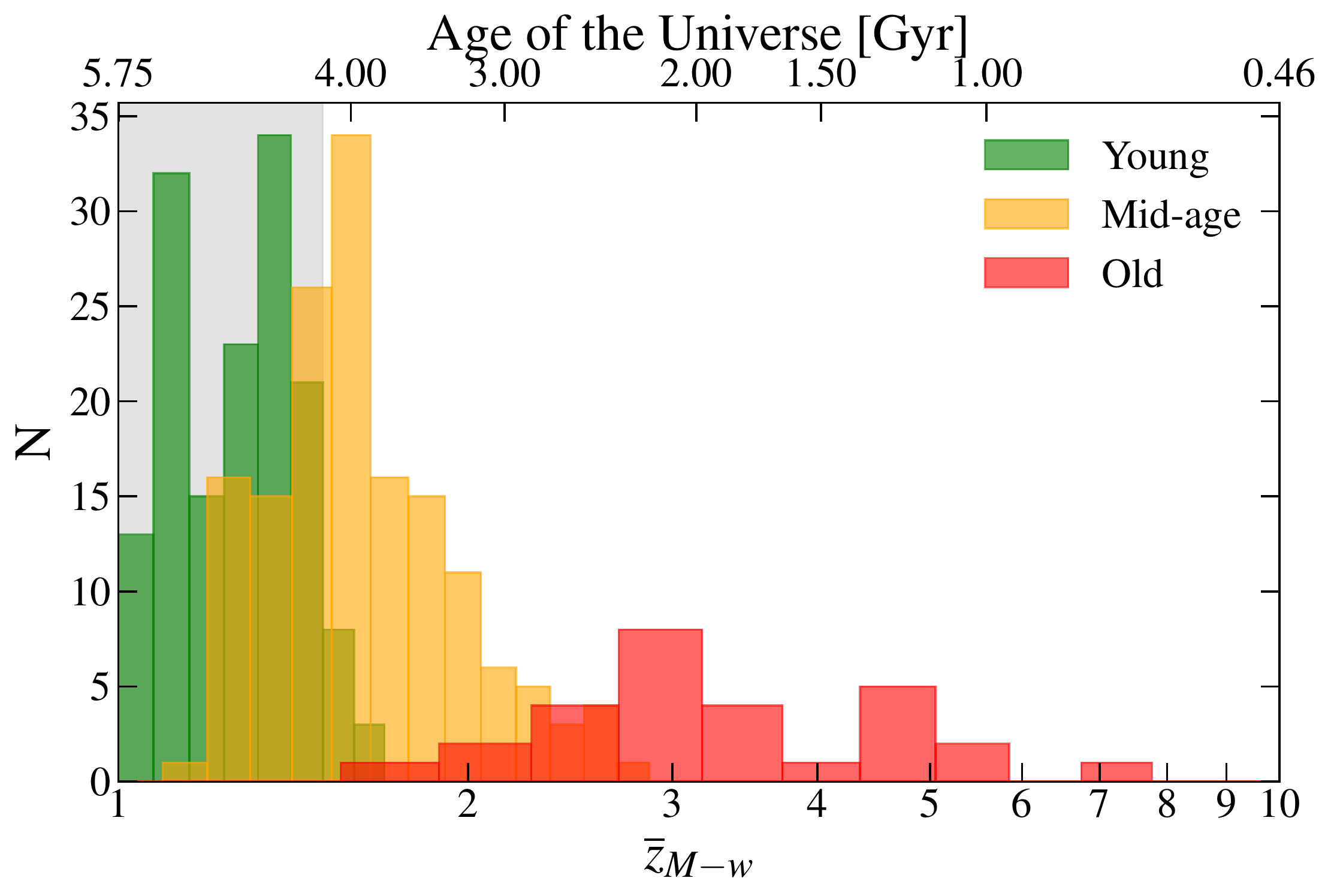} &
    \includegraphics[width=0.521\linewidth, trim= 0 5 5 -5,clip]{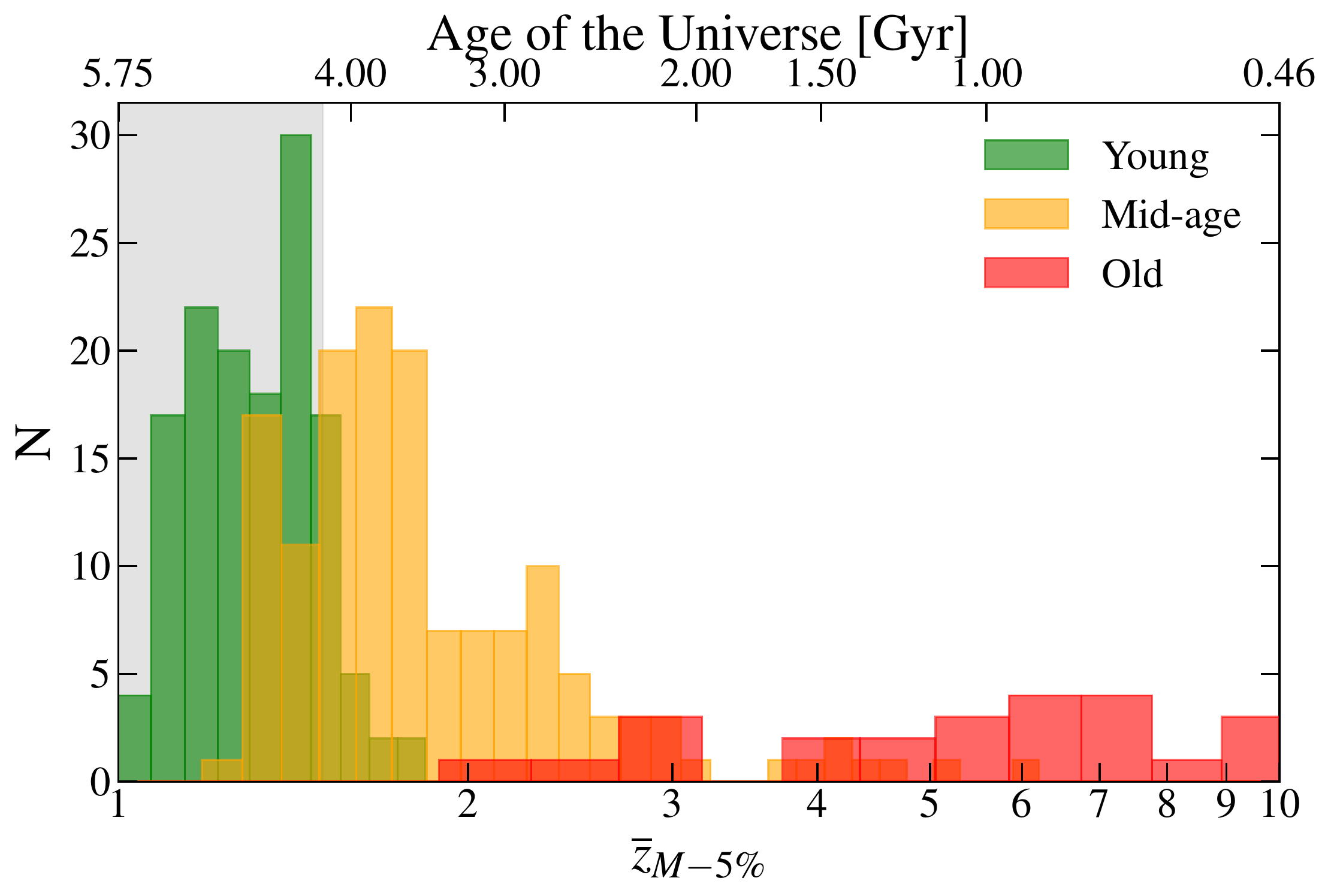}\\
    (a) & (b)
    \end{tabular}
    \caption{Mass-weighted redshift (left panel) and $\overline{z}_{M-5\%}$ (right panel) distribution  of galaxies divided according to their age. Green, orange and red histograms represent young, mid-age, and old galaxies, respectively. The gray area marks the redshift studied in this work, $1.0<z<1.5$. }
    \label{f:zform}
\end{figure*}

\begin{figure}[h]
  \hspace*{-0.2cm}%
    \includegraphics[width=0.5\textwidth]{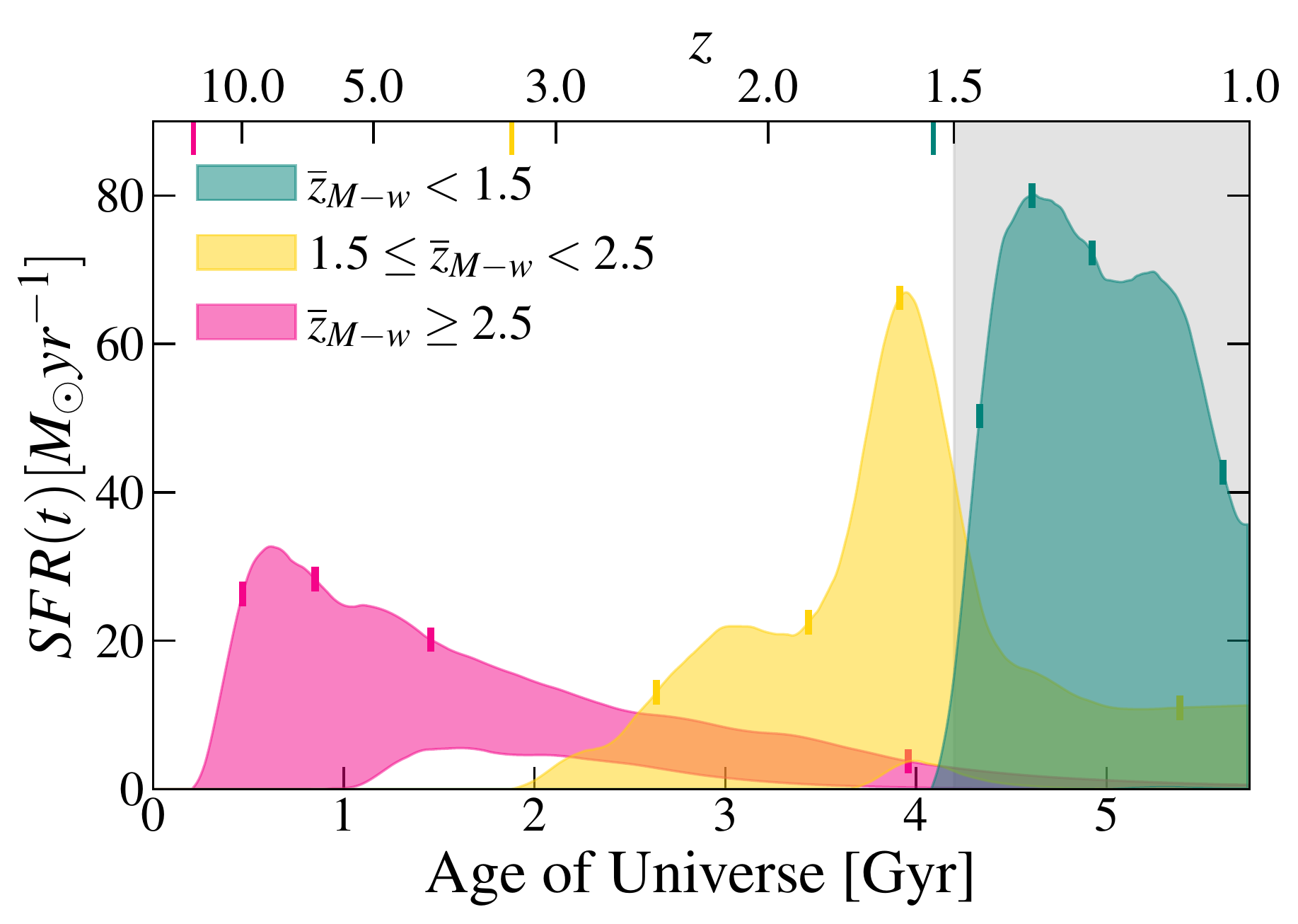}
    \caption{SFHs of galaxies with different mass-weighted formation redshift. The shaded colored areas correspond to the 16th–84th percentile interval computed from the scatter of the SFHs. The gray area marks the redshift studied in this work.The ticks on top on the SFHs represent the redshift at which 5\%, 25\%, 50\% and 95\% of the mass is formed. The colored tick marks in the top x-axis show the redshift at which the SFR start to be greater than 0, for each population.}
    \label{f:average_sf_zform}
\end{figure}

The first class is formed almost entirely (98\%) of star-forming galaxies, if we define these as sources above the MS-3$\sigma$ line of the MS. Moreover, 86\% of the galaxies with $\overline{t}_{M-w}<$0.5~Gyr are above the MS-1$\sigma$ line. The second class is formed by star-forming and passive galaxies in almost equal percentage (47\% and 53\%  star-forming and quiescent, respectively, defining the distinction in the MS-3$\sigma$). The third class is mainly formed by quiescent galaxies (86\%).
Young and mid-age galaxies form the bulk of our sample, 46 and 45 \% respectively, while 9\% of the sample corresponds to old galaxies.
The median mass-weighted age of each of the three classes is $\overline{t}_{M-w}=$0.20$\pm$0.12 Gyr for young galaxies, $\overline{t}_{M-w}=$1.05$\pm$0.40 Gyr for mid-age galaxies, and $\overline{t}_{M-w}=$2.72$\pm$0.45 Gyr for old galaxies. Other statistical properties of the three galaxy populations are shown in Table~\ref{t:properties}, medians together with the 16th and 84th percentiles of their distribution. Young galaxies have higher SFRs, lower masses and shorter formation time-scales, while old galaxies have lower SFRs, slightly higher masses and longer formation time-scales. In terms of dust attenuation, young galaxies are more dusty ($\mathrm{A_V\, =\, 1.8 \pm 0.6}$~mag) than old ones ($\mathrm{A_V\, =\, 0.8 \pm 0.5}$~mag). We note that the most significant difference in the attenuation can be found between young and mid-age galaxies, i.e., the effect of dust disappears rapidly in less than 1~Gyr, and then remains roughly unchanged for several Gyr.
To properly compare galaxies observed at different redshifts, we derived the redshift corresponding to the mass-weighted ages, i.e., the mass-weighted formation redshift $\mathrm{\overline{z}_{M-w}}$ \citep[as done in][]{costantin2021, costantin2022}. 
Fig.~\ref{f:zform}(a) shows the mass-weighted formation redshift of the three classes of galaxies defined above. Fig.~\ref{f:zform}(b) shows the redshift at which the three classes of galaxies had already formed 5\% of their total mass ($\mathrm{\overline{z}_{M-5\%}}$).
We decided to divide our sample also in another three classes based on the galaxy mass-weighted formation redshift:
\begin{itemize}
    \item galaxies with $\overline{z}_{M-w}<$1.5,
    \item galaxies with 1.5$\leq \overline{z}_{M-w}<$2.5,
    \item galaxies with $\overline{z}_{M-w} \geq$2.5.
\end{itemize}  
Table \ref{t:properties_zform} shows the properties of the galaxies in these three groups. Figure~\ref{f:average_sf_zform} shows the median SFHs of galaxies in these three groups. 

\begin{deluxetable}{lcccc}[ht]
	\tablewidth{0pt}
	\tablecaption{50, 16th (lower) and 84th (upper) percentiles of the distribution of galaxy properties for galaxies of different ages. The properties refer to the SED fitting ( $\mathrm{\overline{t}_{M-w}}$, $\mathrm{\overline{z}_{M-w}}$, $\mathrm{z_{M-5\%}}$,
	$\mathrm{\tau}$, $\mathrm{log(M/M_{\odot})}$, $\mathrm{Z}$, $\mathrm{A_V}$, see Sect.\ref{ss:sed_model}),  environment ($n$, $\mathrm{\Sigma}$, $n_3$, see Sect.~\ref{ss:env_def}), and morphology (S\'ersic index, $\mathrm{\mathrm{R_e}}$, see Sect.~\ref{ss:morpho}). 
		\label{t:properties}}
	\tablehead{\colhead{} & \colhead{Young}  & \colhead{Mid-age} & \colhead{Old}}
	\startdata 
	Parameter & & &\\
	\hline
	Number & 152 & 150 & 30 \\
	Number density $\mathrm{[10^{-3} Mpc^{-3}]}$ & $1.03_{0.08}^{0.08}$ & $1.01_{0.08}^{0.08}$ & $0.20_{0.04}^{0.04}$\\
	Fraction & 46\% & 45\% & 9\% \\
	$z$ & $\mathrm{1.22_{1.02}^{1.37}}$ & $\mathrm{1.22_{1.02}^{1.41}}$ & $\mathrm{1.37_{1.15}^{1.45}}$\\ 
	$\mathrm{\overline{t}_{M-w}\, [Gyr]}$ & $\mathrm{0.2_{0.1}^{0.3}}$ & $\mathrm{1.1_{0.7}^{1.5}}$ & $\mathrm{2.7_{2.3}^{3.2}}$\\
	$\mathrm{log(M/M_{\odot})}$ & $\mathrm{10.34^{10.69}_{10.09}}$ & $\mathrm{10.42_{10.14}^{10.75}}$ & $\mathrm{10.53_{10.24}^{10.75}}$ \\
	$\tau$ [Myr] & $\mathrm{127^{100}_{396}}$ & $\mathrm{198^{100}_{626}}$ & $\mathrm{513^{396}_{983}}$ \\
    $\overline{z}_{M-w}$ & $\mathrm{1.32_{1.12}^{1.47}}$ & $\mathrm{1.60_{1.37}^{2.01}}$ & $\mathrm{3.21_{2.43}^{4.99}}$\\
    $\overline{z}_{M-5\%}$ & $\mathrm{1.30_{1.11}^{1.47}}$ & $\mathrm{1.70_{1.37}^{2.40}}$ & $\mathrm{6.20_{3.73}^{9.94}}$\\
	$\mathrm{SFR\, [M_{\odot} yr^{-1}]}$ & $\mathrm{35_{17}^{74}}$ & $\mathrm{5.0_{1}^{39}}$ & 
	$\mathrm{3_{1}^{9}}$\\
	$\mathrm{A_V\, [mag]}$ & $\mathrm{1.8_{1.3}^{2.5}}$ & $\mathrm{0.9_{0.2}^{1.6}}$ & $\mathrm{0.8_{0.3}^{1.0}}$\\
	$\mathrm{Z\, [Z_{\odot}]}$ & $\mathrm{0.4_{0.2}^{1.0}}$ & $\mathrm{1.0_{0.2}^{2.5}}$ & $\mathrm{1.0_{0.2}^{2.5}}$\\
	$n\, \mathrm{[Mpc^{-2}]}$ & $0.95_{0.40}^{2.36}$ & $0.79_{0.32}^{2.60}$ & $0.57_{0.32}^{1.28}$\\
	$\mathrm{\Sigma\, [10^{10} M_{\odot}Mpc^{-2}]}$ & $3.9_{1.0}^{13.3}$ & $3.3_{1.2}^{13.3}$ & $2.8_{0.9}^{8.3}$\\
	$n_3\, \mathrm{[Mpc^{-2}]}$ & $0.19_{0.04}^{1.01}$ & $0.15_{0.03}^{1.36}$ & $0.10_{0.04}^{0.48}$\\
	S\'ersic index $n$ & $\mathrm{1.3_{0.7}^{2.9}}$ & $\mathrm{2.8_{1.4}^{5.0}}$ & $\mathrm{4.0_{2.3}^{5.4}}$\\
    Re [kpc] & $\mathrm{3.86_{2.35}^{6.01}}$ & $\mathrm{2.16_{0.96}^{4.68}}$ & $\mathrm{1.48_{0.72}^{3.27}}$\\
	\enddata
\end{deluxetable}

~
\begin{deluxetable}{lcccc}
	\tablewidth{0pt}
	\tablecaption{50, 16th (lower) and 84th (upper) percentiles of the distribution of galaxy properties (same as in Table~\ref{t:properties}) for galaxies of different mass-weighted formation redshift: $\mathrm{\overline{z}_{M-w}<1.5}$ (Class 1), $\mathrm{1.5 \leq \overline{z}_{M-w}<2.5}$ (Class 2) and $\mathrm{\overline{z}_{M-w}\geq2.5}$ (Class 3).  
		\label{t:properties_zform}}
	\tablehead{\colhead{} & \colhead{Class 1}  & \colhead{Class 2} & \colhead{Class 3}}

	\startdata 
	Parameter & & &\\
	\hline
	Number & 180 & 123 & 29 \\
	Number density $\mathrm{[10^{-3}\, Mpc^{-3}]}$ & $1.26_{0.09}^{0.09}$ & $0.79_{0.07}^{0.07}$ & $0.20_{0.04}^{0.04}$\\
	Fraction & 54\% & 37\% & 9\% \\
	$z$ & $\mathrm{1.14_{1.02}^{1.32}}$ & $\mathrm{1.24_{1.14}^{1.44}}$ & $\mathrm{1.43_{1.27}^{1.47}}$\\ 
	$\mathrm{\overline{t}_{M-w}\, [Gyr]}$ & $\mathrm{0.3_{0.1}^{0.8}}$ & $\mathrm{1.1_{0.5}^{1.6}}$ & $\mathrm{2.7_{1.9}^{3.2}}$\\
	$\mathrm{log(M/M_{\odot})}$ & $\mathrm{10.38_{10.11}^{10.72}}$ & $\mathrm{10.42_{10.14}^{10.73}}$ & $\mathrm{10.54_{10.21}^{10.76}}$ \\
	$\tau$ [Myr] & $\mathrm{126_{100}^{311}}$ & $\mathrm{233_{101}^{726}}$ & $\mathrm{527_{398}^{1005}}$ \\
    $\overline{z}_{M-w}$ & $\mathrm{1.32_{1.13}^{1.44}}$ & $\mathrm{1.66_{1.55}^{2.04}}$ & $\mathrm{3.21_{2.72}^{4.98}}$\\
    $\overline{z}_{M-5\%}$ & $\mathrm{1.43_{1.15}^{1.80}}$ & $\mathrm{3.08_{1.37}^{5.24}}$ & $\mathrm{8.99_{7.18}^{10.90}}$\\
    $\mathrm{SFR\, [M_{\odot} yr^{-1}]}$ & $\mathrm{28_{5}^{62}}$ & $\mathrm{6_{1}^{50}}$ & $\mathrm{5_{1}^{27}}$\\
	$\mathrm{A_V\, [mag]}$ & $\mathrm{1.7_{0.7}^{2.3}}$ & $\mathrm{1.0_{0.3}^{1.7}}$ & $\mathrm{0.8_{0.3}^{1.2}}$\\
	$\mathrm{Z\, [Z_{\odot}]}$ & $\mathrm{0.4_{0.2}^{2.5}}$ & $\mathrm{0.4_{0.2}^{2.5}}$ & $\mathrm{1.0_{0.2}^{2.5}}$\\
	$n\, \mathrm{[Mpc^{-2}]}$ & $0.19_{0.40}^{2.36}$ & $0.15_{0.32}^{2.60}$ & $0.57_{0.32}^{1.28}$\\
	$\mathrm{\Sigma\, [10^{10} M_{\odot}Mpc^{-2}]}$ & $5.0_{1.1}^{24.8}$ & $4.0_{1.1}^{9.6}$ & $3.0_{0.8}^{3.2}$\\
	$n_{3}\, \mathrm{[Mpc^{-2}]}$ & $0.26_{0.05}^{1.72}$ & $0.18_{0.03}^{0.72}$ & $0.08_{0.04}^{0.20}$\\
	S\'ersic index  $n$ & $\mathrm{1.8_{0.7}^{3.9}}$ & $\mathrm{2.6_{1.1}^{4.4}}$ & $\mathrm{3.8_{1.5}^{5.5}}$\\
    Re [kpc] & $\mathrm{3.56_{1.83}^{5.64}}$ & $\mathrm{2.32_{0.97}^{4.97}}$ & $\mathrm{1.42_{0.67}^{3.28}}$\\
	\enddata
\end{deluxetable}

\subsection{Volume Number densities of galaxies in different evolutionary stages} 
\begin{figure}[h]
    \centering
    \includegraphics[scale=0.4]{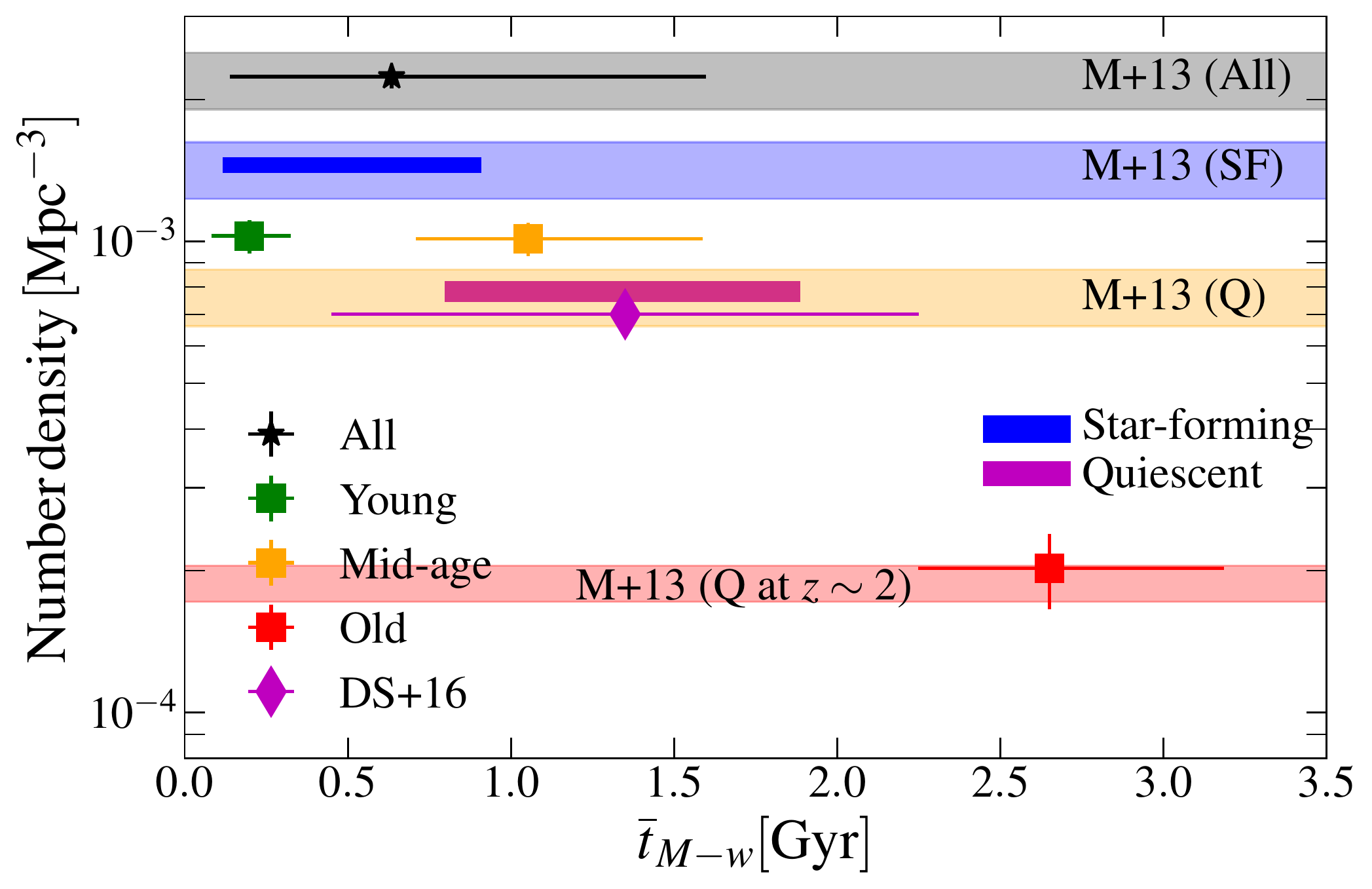}
    \caption{Number densities of different types of massive galaxies at $1.0<z<1.5$, classified by age and star formation activity (distance to the MS). We show
    young (green), mid-age (orange) and old (red) massive galaxies, as well as number densities for star-forming (blue rectangle) and quiescent (red rectangle). The number density of quiescent galaxies from \citetalias{ds2016} is shown as a magenta diamond. The black, blue and orange shaded areas represent the number density of all, star-forming and quiescent galaxies at $1.0 \leq z \leq 1.5$ from \citet{muzzin2013}, respectively. The red shaded area represent the number density of quiescent galaxies at $z\sim2$ \citet{muzzin2013}.}
    \label{f:nd}
\end{figure}

In this section, we derive the volume number densities of different types of galaxies in our sample, and compare them with those derived from other works (\citetalias{ds2016}, \citealt{muzzin2013}). The catalog from which we have selected our sample has a mass completeness limit of $\mathrm{10^{9.5}M_{\odot}}$ up to $z=3$. We derive the number density of massive galaxies, both classifying them by age (as defined in Section~\ref{ss:tM}), and star-formation activity. Then we check for IR emitters with significant fluxes at 24 $\mu$m, and we remove them from the quiescent sample. Since the SFRs for the galaxies in our sample comes from IR and UV indicators, and it is not a direct product of the SED fitting, the two classifications are independent to some extent, i.e., the former use only 2-3 points in the UV part of the SED, jointly with 1-6 point in the mid- and far-IR, the latter use all SHARDS and grism data points (on average, 120 independent fluxes).\\
The volume number densities are derived using the $1/V_{max}$ technique \citep{schmidt1968}, while their errors are Poisson uncertainties.
In Figure~\ref{f:nd}, we show the volume number density of the entire sample, young, mid-age, and old galaxies, and star-forming and quiescent galaxies. We compare these values with the number densities of galaxies in \citetalias{ds2016}, and \cite{muzzin2013}. In \cite{ds2016}, they compute the number densities of massive ($\mathrm{>10^{10}M_{\odot}}$), quiescent galaxies at 1.0$<$$z$$<$1.5 in this same GOODS-N field, while in \cite{muzzin2013} they compute the number densities for the sample of all, star-forming, and quiescent galaxies in the COSMOS field.
The number density of our sample of massive galaxies is: $n\, =\, 2.15\, \pm\, 0.12 \times 10^{-4}~\mathrm{Mpc^{-3}}$. This is in excellent agreement within 1$\sigma$) with the number densities of galaxies above $10^{10}M_{\odot}$ reported in \cite{muzzin2013} at $z = 1.0-1.5$ ($n\, = 2.19 \pm 0.06 \times 10^{-3}~\mathrm{Mpc^{−3}}$). The number density of quiescent galaxies is:  $n\, =\, 7.4\, \pm\, 0.7 \times 10^{-4} \mathrm{Mpc^{-3}}$, in good agreement with the values found in \citetalias{ds2016}, $n\, =\, 7.0\, \pm\, 0.7 \times 10^{-4}~\mathrm{Mpc^{-3}}$, and in \cite{muzzin2013}, $n\, =\, 7.6\, \pm\, 0.2 \times 10^{-4} ~\mathrm{Mpc^{-3}}$. The number density of star-forming galaxies is: $n\, = 1.4 \pm 0.1 \times 10^{-3}~\mathrm{Mpc^{−3}}$, in agreement with the values reported in \citep{muzzin2013}, $n\, = 1.41 \pm 0.04 \times 10^{-3}~\mathrm{Mpc^{−3}}$.

Our results point out that at $1.0 \leq z \leq 1.5$ we have roughly the same volume density of galaxies up to 1.5~Gyr old, while $> 2$~Gyr old galaxies represent around 10\% of the whole galaxy population. 
The number density of our old galaxies at $1.0 \leq z \leq 1.5$ is completely consistent with the number density of quiescent galaxies at $z \sim 2$ (\citealt{muzzin2013}), which invites to identify the latter as progenitors of the former. Globally speaking about the population of massive galaxies at $z > 1$, we can say that although some small number of massive galaxies start to form very early in the history of the Universe, beyond $z \sim 2.5$, the bulk of the massive galaxy population starts to appear at $z \sim 2$. 

\section{Environment}\label{s: env}

In this section, we will discuss the interplay between galaxy assembly histories and environment. 

\subsection{Environment estimations}\label{ss:env_def}

Here we describe the different methods that we used to estimate the local environment around a given source, benefiting from the unprecedentedly accurate SHARDS+CANDELS photometric redshifts available for GOODS-N galaxies \citep{barro2019}. The environment of a galaxy is typically estimated through the density of galaxies located in its immediate proximity. There are a variety of possible density measures that can be employed for this purpose \citep{cooper2005, muldrew2012}. In this work, we used three different methods:

\begin{itemize}
    \setlength\itemsep{-0.2em}
    \item number density within a fixed aperture, \textit{n}, in units of Mpc$^{-2}$;
    \item surface mass density, $\mathrm{\Sigma}$, in units of M$_\odot$\,Mpc$^{-2}$;
    \item number density based on the distance to the Nth nearest neighbor {$n^{Nth}$}, also  in units of Mpc$^{-2}$.
\end{itemize}

\noindent {\it {Number density within a fixed aperture}.} This method probes a volume around each galaxy, within which the number of neighbours are counted. We estimated galaxy environment as the number of galaxies inside a cylinder with base radius R and height $\Delta z$, centered on each galaxy of the sample, using all  galaxies above a given mass limit as tracers. For the redshift of interest, $1.0<z<1.5$, after several tests, we decided to use a radius of R = 1 Mpc. The height of the cylinder ($\Delta z$) is proportional to the error on the photometric redshift:

\begin{equation}
    \Delta z=n \sigma_{ \Delta z/(1+z)} (1+z).
\end{equation}
where $n=1.5$ and $\sigma_{\Delta z/(1+z)}$ = 0.0028 for the entire SHARDS+CANDELS sample, as defined in Section~\ref{s:sample_sel}. The value of $\sigma_{\Delta z/(1+z)}$ is the same when calculated on the entire catalog, or if we restrict the calculation at the redshift of interest 1.0$<$$z$$<$1.5. A high value of $\Delta z$ increases the accuracy in dense environments, but at the same time sacrifices sensitivity at low densities. Our choice of $n$, and consequently of $\Delta z$, is best suited for a broad range of environments \citep{krefting2020}.

\noindent {\it {Surface mass density}}. This method is a modification of the fixed aperture, where instead of simply counting the galaxies in a cylinder of given area and height, we consider the total mass within that given cylinder.

\noindent {\it  Density based on the distance to the Nth nearest neighbor ($n_{Nth}$)}. 
The surface density of a galaxy is expressed as: $n_{Nth}=N/( \pi d_N^{2})$, where N is the number of galaxies and $d_N$ is the angular diameter distance to the Nth-nearest neighbor. The most common choices for N are: N=3, 5. Here, we used N=3. \\
In the three different environmental estimations, only galaxies within $\Delta z\, =\, n \sigma_{ \Delta z/(1+z)} (1+z)$ of a given source enter in the calculation. To calculate both number and surface mass density, the target galaxy is also included.\\
We repeated our analysis considering only galaxies within 1 Mpc from the border of our FoV, to highlight possible edge effects, finding that our results would not change.

\subsection{Environment of massive galaxies}{\label{ss: env_res}}

\begin{figure*}[ht]
\begin{tabular}{cc}
  \hspace*{-1cm}%
  \includegraphics[width=90mm]{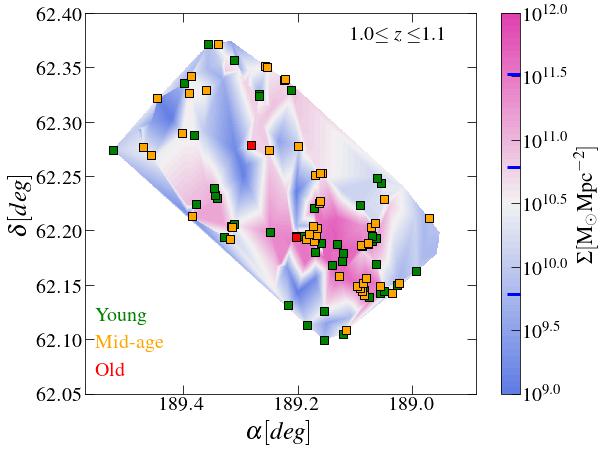} &  \includegraphics[width=90mm]{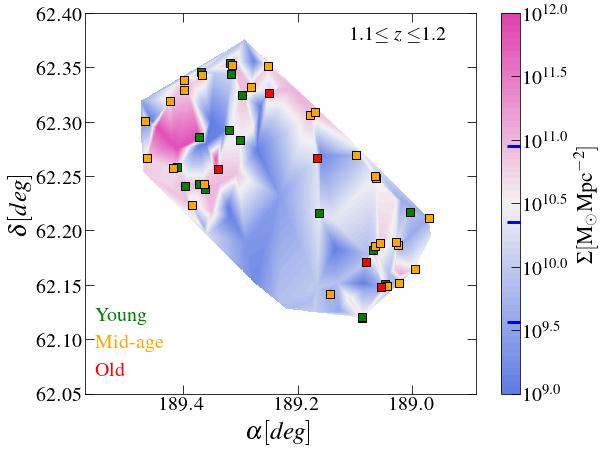} \\
    \hspace*{-1cm}%
(a) & (b) \\[6pt]
  \hspace*{-1cm}%
 \includegraphics[width=90mm]{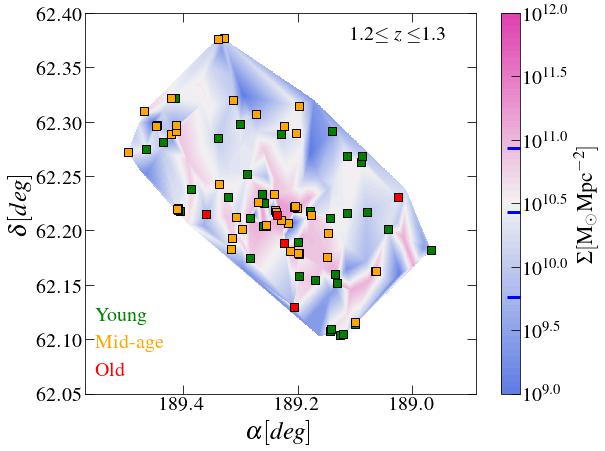} &   \includegraphics[width=90mm]{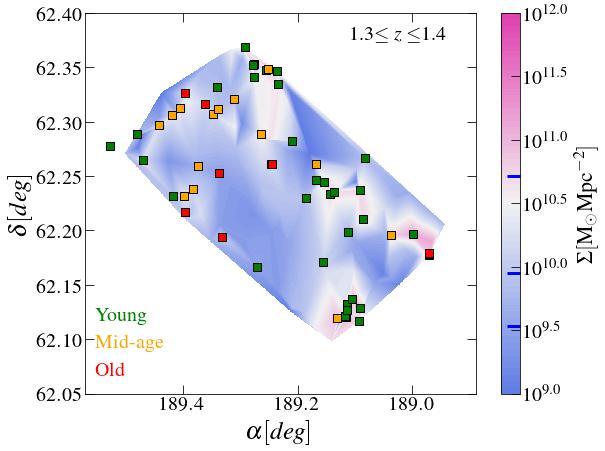} \\
(c) & (d) \\[6pt]
 \multicolumn{2}{c}{
 \includegraphics[width=90mm]{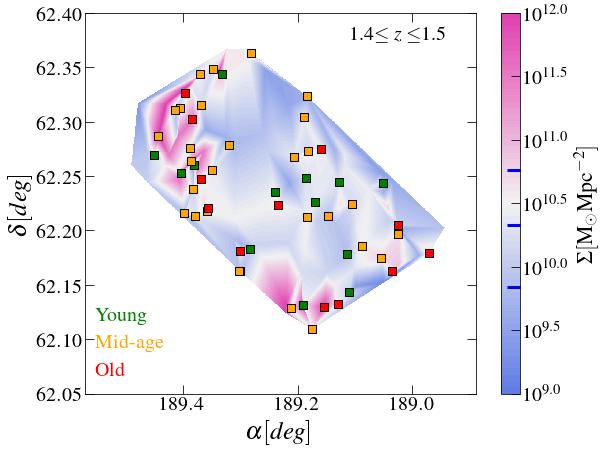} } \\
\multicolumn{2}{c}{(e)}
\end{tabular}
\caption{Mass surface density maps in different redshift bins, constructed to enclose approximately the same number of galaxies and with a width typically 10 times larger than the $\mathrm{\sigma_{nmad}}$ photometric redshift accuracy. We overplot the positions of young (green), mid-age (orange), and old (red) galaxies. Blue lines in the colorbars represent the 16th, 50th, 84th percentiles of the surface mass density of all galaxies with masses $\mathrm{10^{9.5}M_{\odot}}$ in that redshift interval.}
\label{f:ms_maps}
\end{figure*}
Figure~\ref{f:ms_maps} shows the 2D surface mass density maps of galaxies in the GOODS-N field in 5 redshift slices within 1.0$<$$z$$<$1.5. 
The surface mass density of each galaxies has been derived as explained in Section~\ref{ss:env_def}. Here we only considered as tracers galaxies above the mass limit of $10^{9.5}M_{\odot}$, which is the completeness limit of our data at $z$=1.5. Then, a continuous map was obtained by interpolating using the nearest neighbor interpolation method.
According to Figure~\ref{f:ms_maps}, most massive galaxies within our full redshift range are located in regions with densities $\Sigma > 10^{10.2}M_{\odot}\,Mpc^{-2}$. In fact, from Fig.~\ref{f:ms_maps}, it can be seen that, while galaxies above $10^{10}M_{\odot}$, by construction, cannot occupy regions below $\Sigma = 10^{9.5}M_{\odot}\,Mpc^{-2}$ ($\Sigma = \sum_{i} M_{i}/\pi R^2$), they also tend  to avoid regions with densities below $\Sigma = 10^{10}M_{\odot}\,Mpc^{-2}$. Over the entire redshift range ($1\leq z \leq 1.5$), only 7\% of old galaxies are located in dense environment above $\Sigma = 10^{11}M_{\odot}\,Mpc^{-2}$.  
At $1\leq z \leq 1.5$, the median surface mass density for all galaxies above $\mathrm{10^{9.5}M_{\odot}}$ is $\Sigma = \mathrm{2.3_{0.5}^{11.0}\times 10^{10}M_{\odot}Mpc^{-2}}$, while the median surface mass density of our sample of massive galaxies is: $\Sigma = \mathrm{3.3_{1.1}^{13.2}\times 10^{10}M_{\odot}Mpc^{-2}}$. 
The median surface mass density of young massive galaxies is: $\Sigma = \mathrm{3.9_{1}^{13.3}\times 10^{10}M_{\odot}Mpc^{-2}}$, while for mid-age galaxies is: $\Sigma = \mathrm{3.3_{1.2}^{13.3}\times 10^{10}M_{\odot}Mpc^{-2}}$. The median surface mass density of old galaxies is: $\Sigma = \mathrm{2.8_{0.9}^{8.3}\times 10^{10}M_{\odot}Mpc^{-2}}$. The same trend between environment and mass-weighted age is seen when we use the other two environment estimators (see also Table~\ref{t:properties}). 

\begin{figure*}[t]
\begin{tabular}{cc}
  \hspace*{-1.1cm}%
 \includegraphics[width=0.5\linewidth]{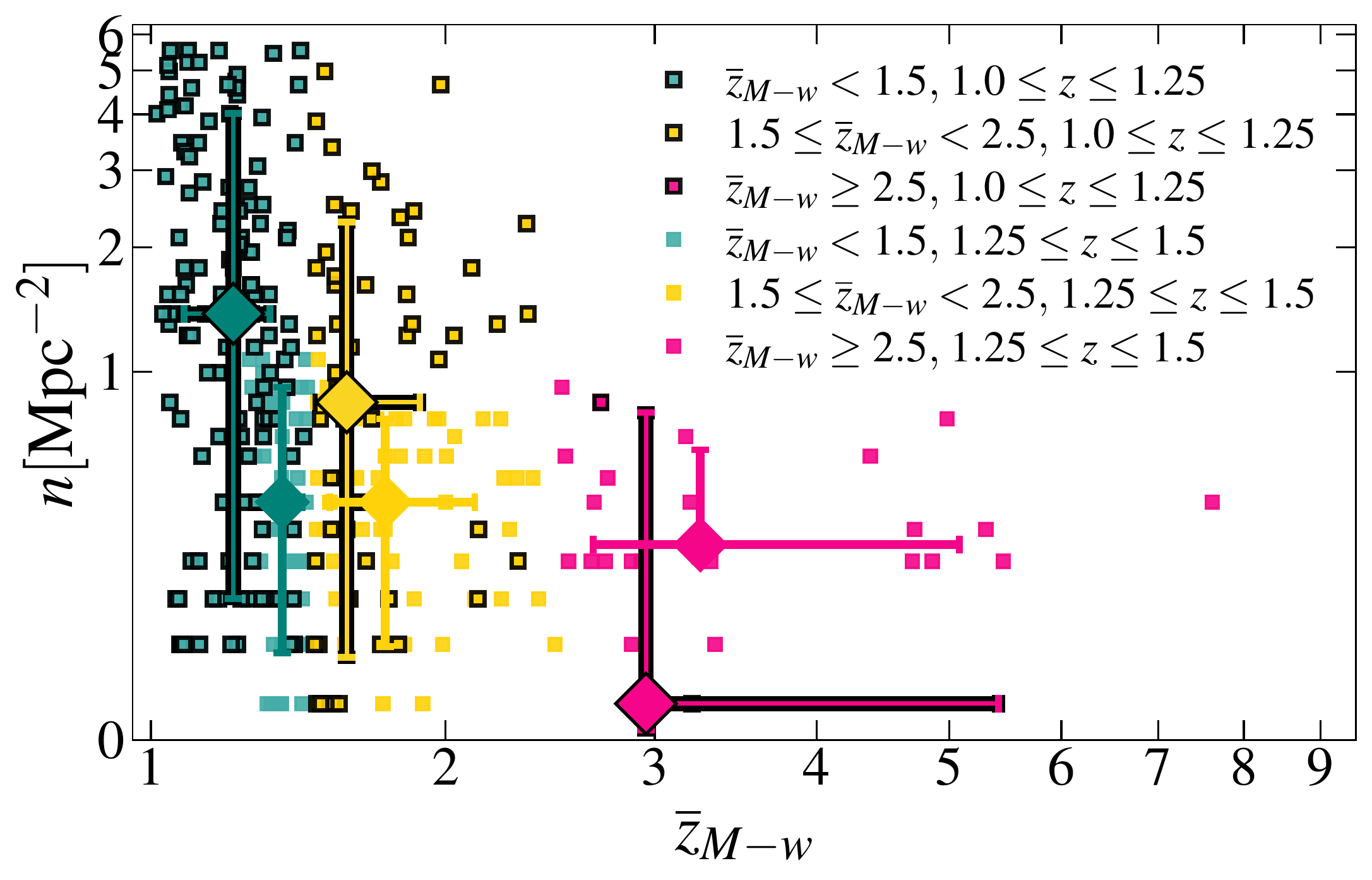} &
 \includegraphics[width=0.52\linewidth]{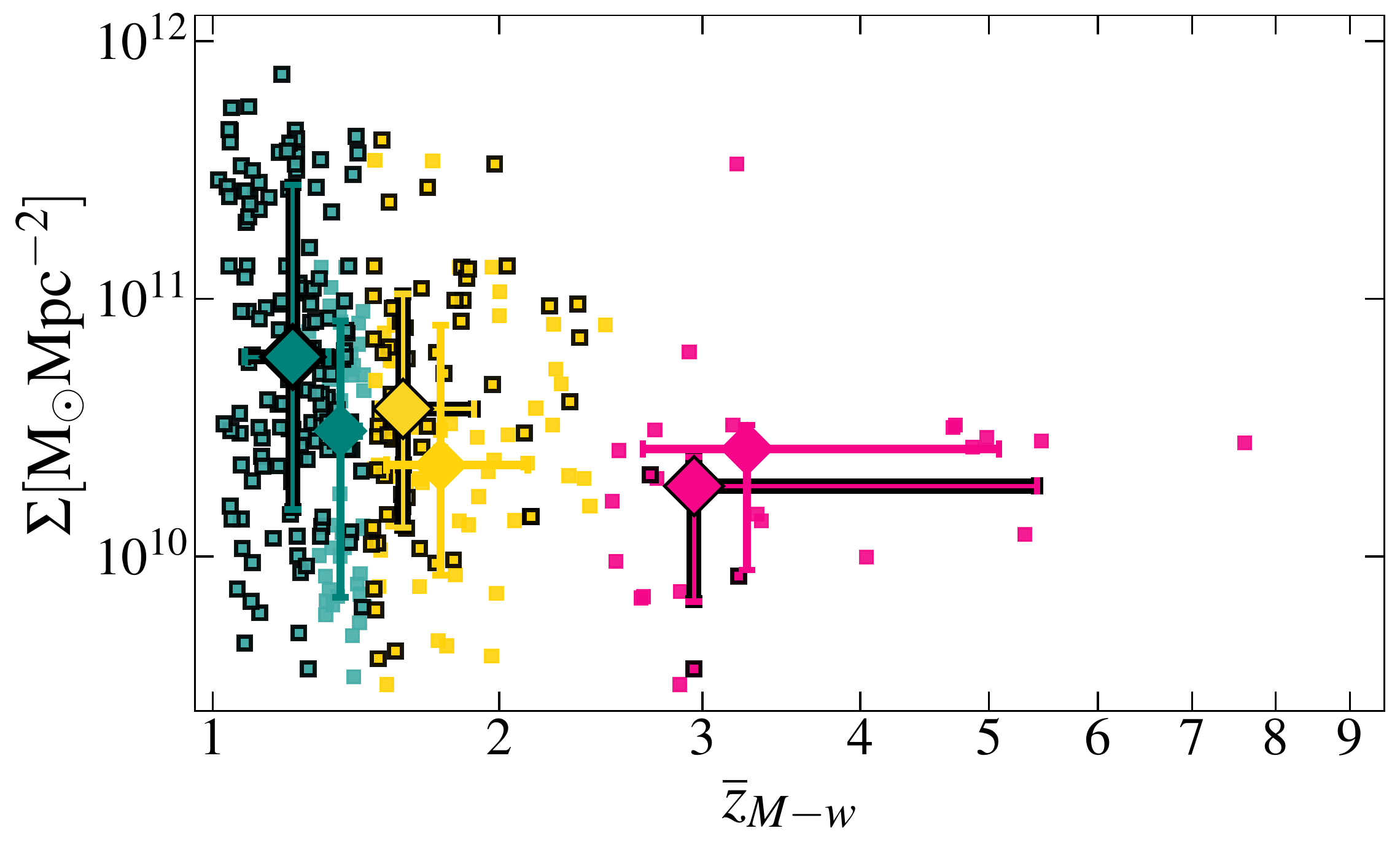} \\
 (a) & (b) \\
 \multicolumn{2}{c}{ \includegraphics[width=0.535\linewidth]{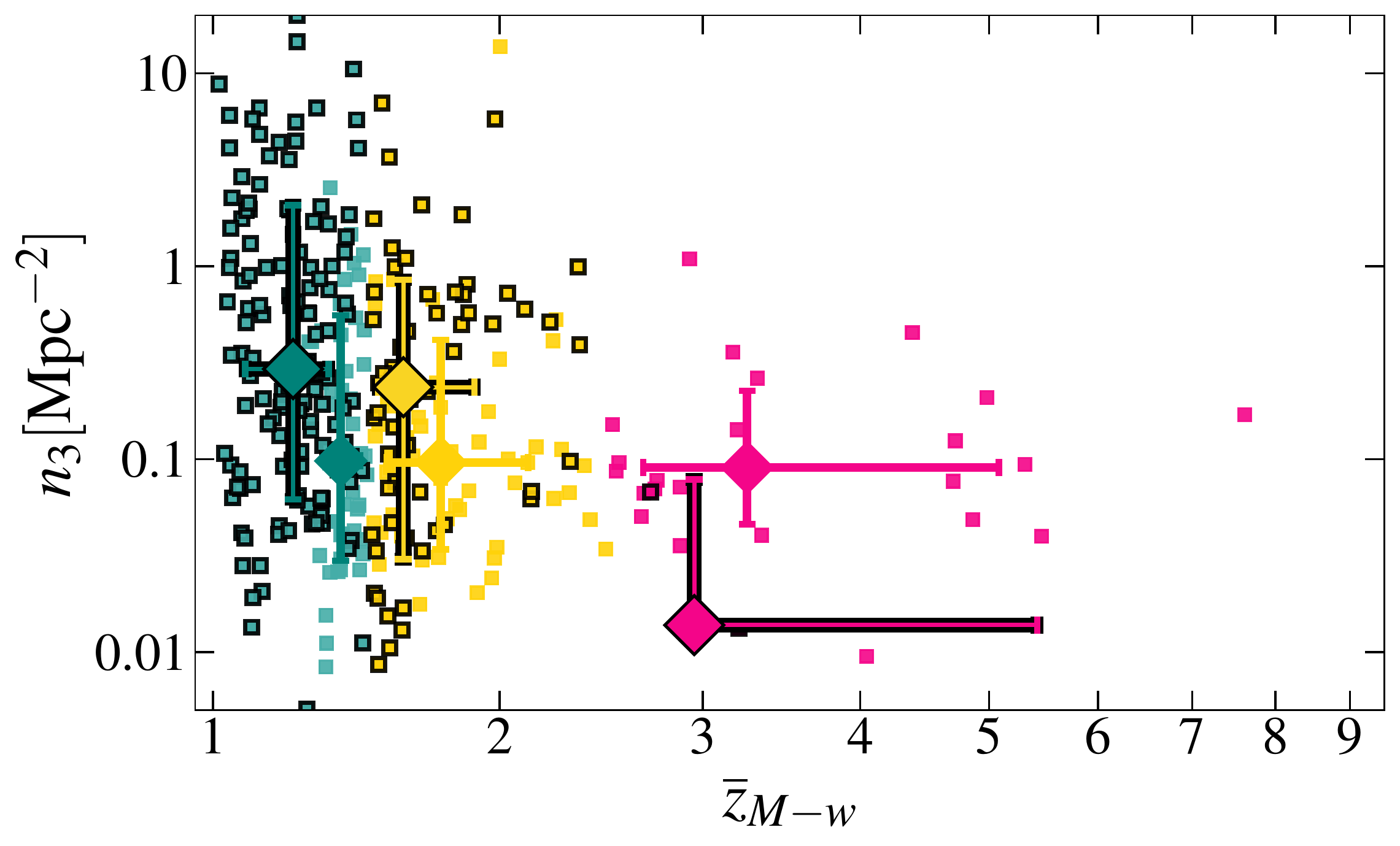}} \\
 \multicolumn{2}{c}{ (c)}
\end{tabular}
\caption{Environment as a function of mass-weighted formation redshift. The environment was estimated using number density within fixed aperture (a), surface mass density (b), and surface mass density based on distance to the third nearest neighbor (c). Galaxies with observed redshift $\mathrm{1.0\leq z < 1.25}$ are marked in black. Light green, yellow and magenta points refer to galaxies in the three classes defined in Table\ref{t:properties_zform}.}
\label{f:estimators}
\end{figure*}

To analyze more quantitatively the environmental dependence of the formation of massive galaxies and the relationship with the SFHs, in Figure~\ref{f:estimators} we show environment vs. mass-weighted formation redshift. The environment was derived by using the three methods defined above, i.e., number density within a fixed aperture (top row), mass surface density (middle row), and distance from the 3$^\mathrm{rd}$ nearest neighbor (bottom row). We show environment vs mass-weighted formation redshift for the three groups of galaxies in Table \ref{t:properties_zform}. Figure~\ref{f:estimators} shows that all three environment estimators lead to similar results, so for the rest of the paper we concentrate on discussing what we learn from the mass surface density.
The highest density environments, i.e., $\mathrm{\Sigma \, >\, 10^{11}\, M_{\odot}\, Mpc^{-2}}$, are predominantly occupied by recently formed galaxies with $\mathrm{\overline{z}_{M-w}<2.5}$. Over the entire redshift range $\mathrm{1.0\, <\, z\, <\, 1.5}$, in the regions with surface mass density of $\mathrm{\Sigma \, >\, 10^{11}\, M_{\odot}\, Mpc^{-2}}$, there are 23\% of the galaxies with $\overline{z}_{M-w}<$ 1.5, 8\% of the galaxies with 1.5$\leq \overline{z}_{M-w} \leq$ 2.5, while only 3\% of the first formed galaxies ($\overline{z}_{M-w} \geq $ 2.5). When we split the redshift range in two, we find that at $\mathrm{1.0\, <\, z\, <\, 1.25}$, 22\% of the galaxies with $\overline{z}_{M-w}<$ 1.5 are found in the overdense environments, while there are no galaxies with $\overline{z}_{M-w}\geq$ 2.5. At higher redshift, $\mathrm{1.25\, <\, z\, <\, 1.5}$, the difference is less pronounced, with $~$18\% of recently formed galaxies ( $\overline{z}_{M-w}<$ 1.5) being in highest density regions, against 4\% of galaxies with $\overline{z}_{M-w}\geq$ 2.5. A Kolmogorov-Smirnov (KS) test was also performed between the distribution of the surface mass density of galaxies in different classes. Both over the entire redshift range and in the two sub-intervals, a low p-value, $P_{KS}<1.0\%$ was found when comparing the surface mass density of massive galaxies with $\overline{z}_{M-w}>2.5$, and galaxies with lower formation redshifts. This means that we can rule out the null hypothesis that those galaxies are drawn from the same parent distribution. 
We repeated our analysis considering as tracers in the different density estimators galaxies above different mass limits, $\mathrm{10^{9.8},\, 10^{10}\, and\, 10^{10.2}\,M_{\odot}}$. The differences in the results are negligible.\\
We also checked the number of neighbors within 1 Mpc and above the mass completeness limit of our sample ($\mathrm{10^{9.5}M_{\odot}}$) that massive galaxies have. At lower redshift, $\mathrm{1.0\, <\, z\, <\, 1.25}$, galaxies with lower formation redshift ($\mathrm{\overline{z}_{M-w}<2.5}$) have, on average, 5 neighbors, with a maximum of 42. Moreover, a significant fraction (20\%) of them has more then 10 neighbors. On the other hand, galaxies formed earlier tend to have less neighbors, with an average of 3, and a maximum of 7. At higher redshift, $\mathrm{1.25\, <\, z\, <\, 1.5}$, the difference between different types of galaxies disappears and all massive galaxies have on average 3 neighbors, with a maximum of 10 neighbors. In summary, at the lowest redshifts, younger galaxies have more neighbors and more mass around, while older galaxies live in less dense environments both in terms of number of neighbor galaxies and their added stellar mass.

\begin{figure}[h]
  \hspace*{-0.1cm}%
    \includegraphics[width=0.5\textwidth]{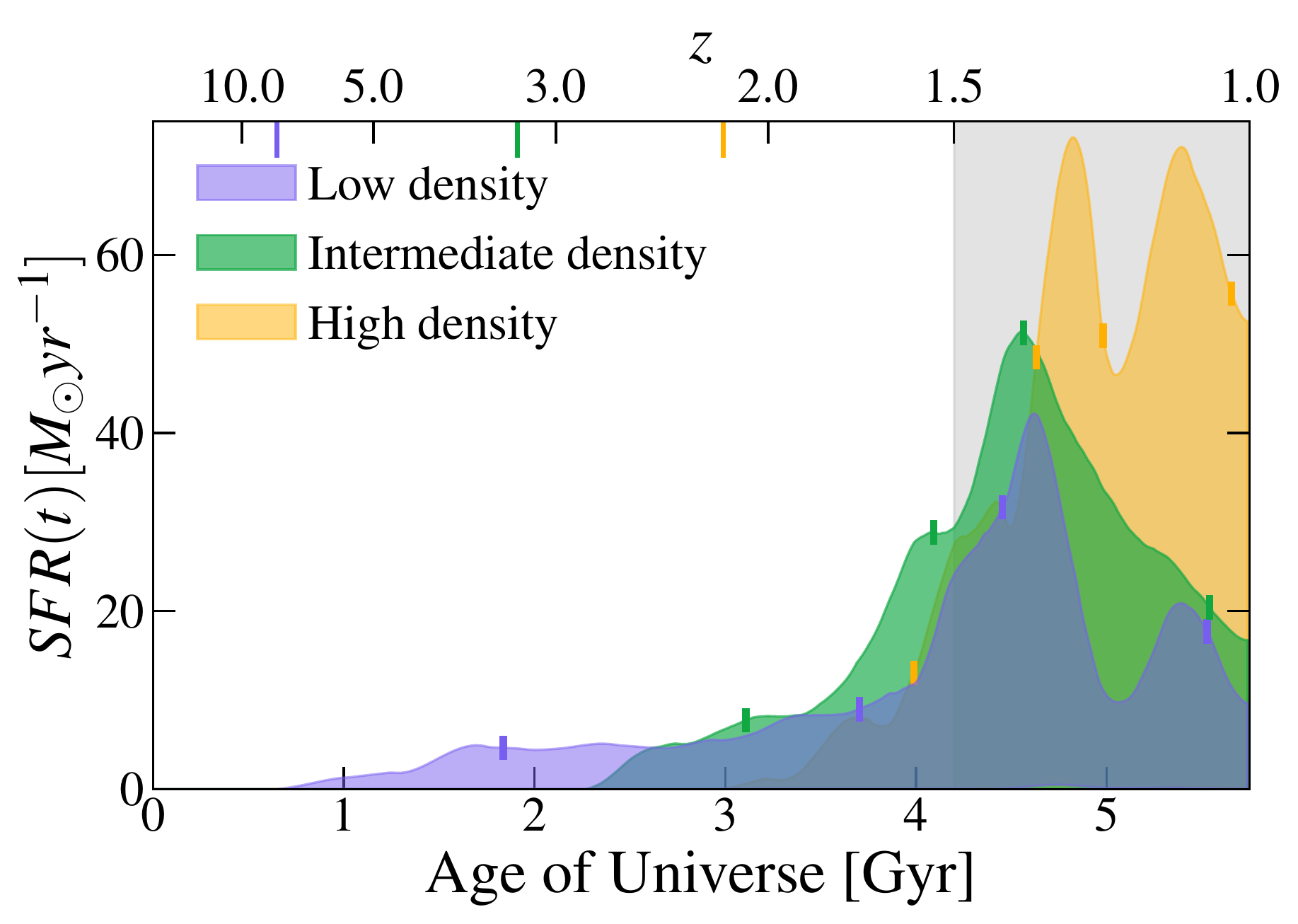}
    \caption{SFHs of galaxies in the lowest density region (purple), at intermediate densities (green) and in the highest density quartile (orange). The format is the same as in Figure~\ref{f:average_sf_zform}. }
    \label{f:average_sf}
\end{figure}

\begin{figure}
  \hspace*{-0.2cm}%
    \includegraphics[width=0.5\textwidth]{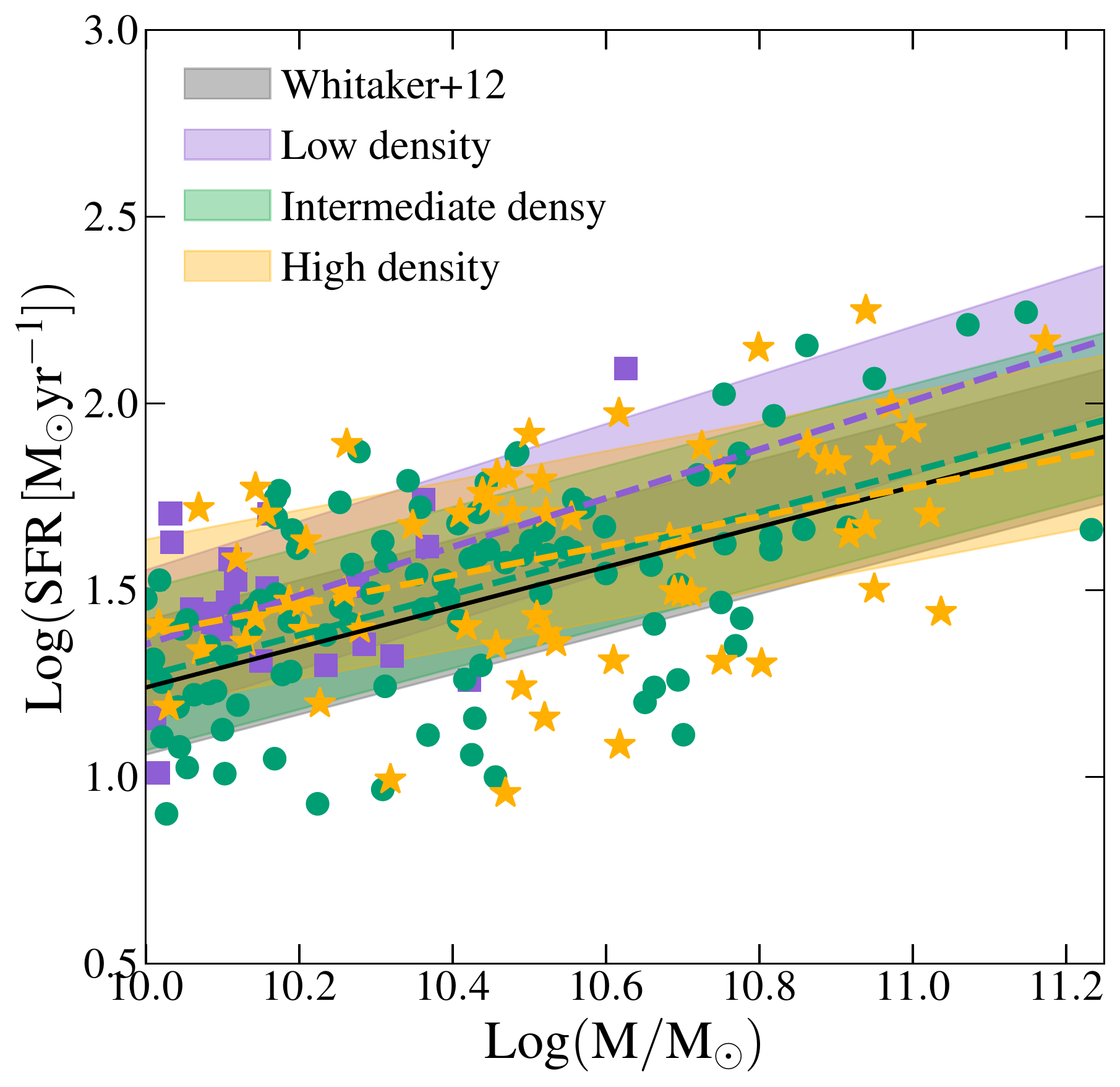}
    \caption{SFR-mass relation for galaxies in the same environments as in Figure~\ref{f:average_sf}. The gray solid line represents the MS in \citet{whitaker2012}, while the gray shaded are correspond to the 1$\sigma$ error quoted in the paper. Dashed lines represent the best-fit MS obtained for galaxies in the three environments. Shaded areas are the 1$\sigma$ errors obtained from the fit. }
    \label{f:ms_env}
\end{figure}

Another way to investigate the interplay between environment and SFHs of massive galaxies is to look into averaged SFHs of galaxies in different environments. To do this, we divided our sample in three different environments, according to the 16th and 84th percentiles of the surface mass density. Figure~\ref{f:average_sf} shows the median SFHs of galaxies in the lowest density quartile ($\mathrm{\Sigma < 10^{10.03}\, M_{\odot}\,Mpc^{-2}}$), at intermediate densities ($\mathrm{10^{10} \leq\, \Sigma\, \leq 10^{11}\, M_{\odot}\,Mpc^{-2}}$), and in the highest density quartile ($\mathrm{\Sigma > 10^{11}\, M_{\odot}\,Mpc^{-2}}$). The SFHs of the galaxies in the three environments peak at similar epochs, around the age of the Universe of 4.5~Gyr, $z=1.3-1.4$, close to the upper limit of our redshift range. This can be expected as the three subsamples are all dominated by young galaxies, present in greater numbers. The SFHs of galaxies in the lowest density quartile start at earlier times (with $z_{M-5\%} \sim 3.5$) and peak at lower SFR values compared to galaxies in denser environments. Indeed, galaxies in the highest density quartile do not present SFHs extending beyond $z\sim2$ and present stronger peak SFRs. Overall, the results in Figures~\ref{f:estimators} and \ref{f:average_sf} are consistent: galaxies that assembled more mass at earlier epochs are caught in relatively less dense environments at $1.0<z<1.5$, and the denser environments are occupied by more recently formed systems, which are also more violently forming stars.
We note that the typical stellar mass of galaxies in the different environments shown in Figure~\ref{f:average_sf} are different. The median stellar mass for the lowest densities is $\mathrm{log(M/M_{\odot})} = 10.15_{10.06}^{10.36}$, $\mathrm{log(M/M_{\odot})} = 10.43_{10.10}^{10.75}$ for the intermediate densities, and $\mathrm{log(M/M_{\odot}) = 10.48_{10.19}^{10.89}}$ for the highest densities.  \\

\begin{figure*}
\centering
 \includegraphics[width=0.7\textwidth]{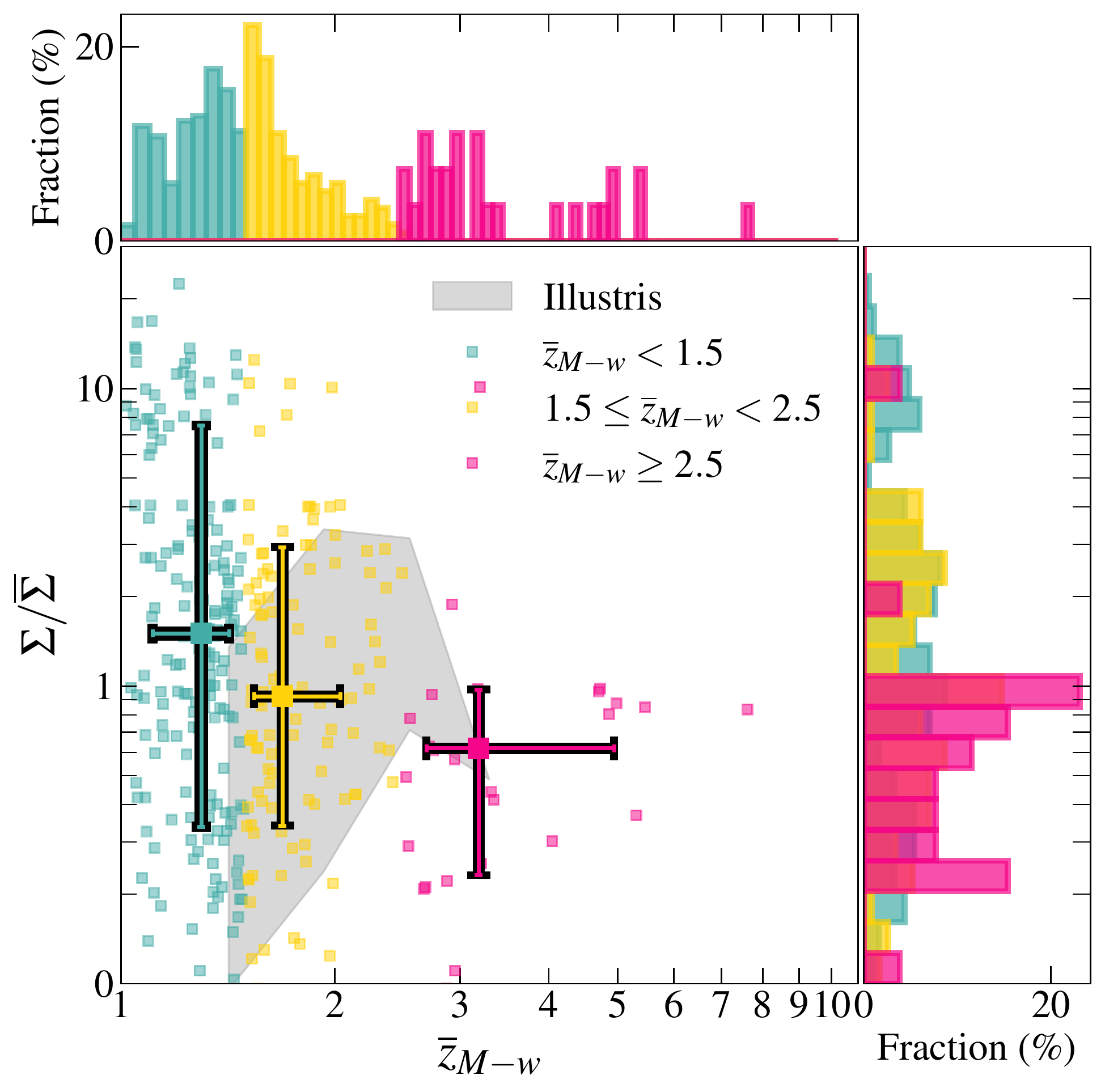}
 \caption{Environment versus mass-weighted formation redshift for our observed massive galaxies in the SHARDS/CANDELS field and for Illustris analogs. The environment is expressed as the ratio between the surface mass density of simulated and observed galaxies and their corresponding mean values, only considering sources above $\mathrm{10^{9.5}M_{\odot}}$. The grey area contains 95\% of all galaxies in the Illustris simulation. Median values and quartiles for real galaxies are shown with colors corresponding to the different types of galaxies presented in Section~\ref{ss:tM}.}
\label{f:ill_comp}
\end{figure*}

We also investigated whether the environment can affect the star formation efficiency of massive galaxies, by comparing the SFR - mass relation of galaxies in the three environments defined above. In Figure~\ref{f:ms_env}, we show the SFR-mass relations of galaxies in the three environments. Here, we focus only on galaxies strictly selected as star-forming, i.e. within 3$\sigma$ from the observed MS in this redshift range \citep{whitaker2012}, excluding starburst galaxies (4\% of our sample). Then, in each environment, we perform again the fit of the MS (shaded areas in Figure~\ref{f:ms_env}). The fit of the MS in the three regions are compatible within the scatter. At the galaxy size scale, environment does not affect the star-formation vs stellar mass relationship. If something, lower density environment present higher SFRs for a fixed stellar mass, maybe revealing a suppression of the star formation linked to neighbor density. Note that this plot show current SFRs, which is different from peak SFRs in the SFHs discussed in the previous paragraphs and for which we indeed found an environmental effect.

\subsection{Comparison with simulations}\label{ss:env_comp}

To better understand the results presented in the previous section, we compare our observational measurements with similar measurements from hydrodynamical simulations.
To perform this comparison, we decided to use the  synthetic deep survey images from the Illustris Project cosmological simulations of galaxy formation \citep{snyder2017}. The Illustris Project consists of hydrodynamical simulations of galaxy formation in a volume 106.5 Mpc across, with detail resolved down to sub-kpc scales. Using the Arepo code \citep{weinberger2020}, Illustris applied galaxy physics consisting of cooling, star formation, gas recycling, metal enrichment, supermassive black hole growth, and gas heating by feedback from supernovae and black holes \citep{vogelsberger2014, nelson2015}. 
The images used for this comparison are the "mock ultra deep fields", images of 2.8 arcmin across, in HST wide filters.
From these images, we selected sources at $1.0<z<1.5$ with masses above $\mathrm{10^{9.5}M_{\odot}}$, and derived the mass-weighted ages of each galaxy by considering its SFH,  which is built from the individual stellar particles' information. We considered stellar particles inside 2 times the galaxy half-mass radius. 
We tested different selections, by considering star particles inside one half-mass radius, or considering all star particles associated to a galaxy. The results we obtained are equivalent, hence we show those obtained for the first case.

\begin{figure*}
\centering
 \includegraphics[width=0.7\textwidth]{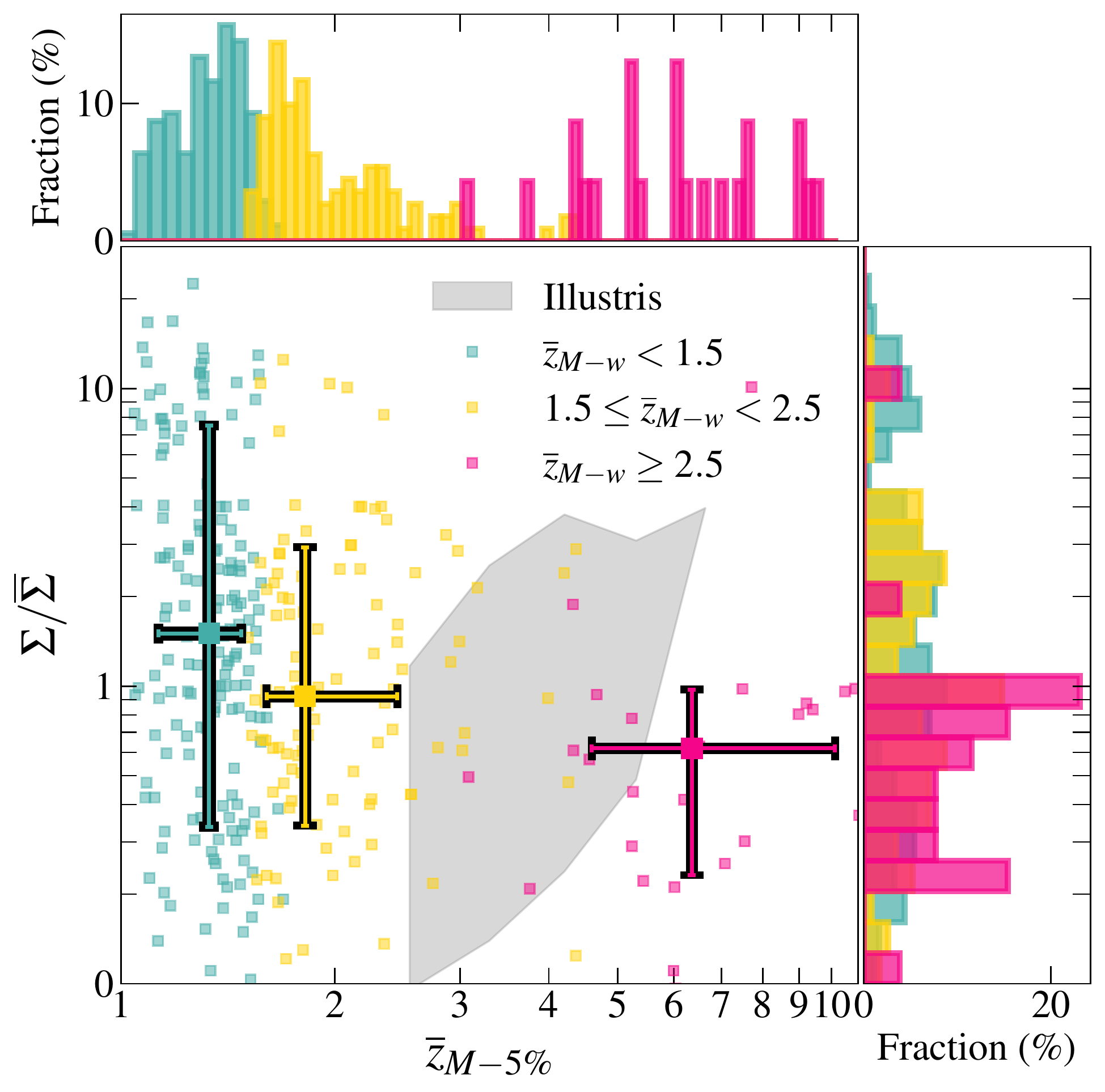}
 \caption{Environment versus,  for our observed massive galaxies in the SHARDS/CANDELS field and for Illustris analogs. The environment is expressed as the ratio between the surface mass density of simulated and observed galaxies and their corresponding mean values, only considering sources above $\mathrm{10^{9.5}M_{\odot}}$. The grey area contains 95\% of all galaxies in the Illustris simulation. Median values and quartiles for real galaxies are shown with colors corresponding to the different types of galaxies presented in Section~\ref{ss:tM}.}
\label{f:ill_comp_z05}
\end{figure*}

The Illustris galaxies in the redshift range studied here ($1<z<1.5$) follow a different MS from the observed one, as already discussed in \cite{sparre2015}. However, when comparing the distance to the Illustris MS with stellar population ages from the simulations, we find a very similar relationship to the one we obtain in GOODS-N for the same mass cut,  $\mathrm{10^{9.5}M_{\odot}}$. As expected considering the small field of view of the simulation, there are not many galaxies with $M>10^{10.5}M_{\odot}$. 
We derived the surface mass density of galaxies in the Illustris simulation by applying the same methods that we used for real galaxies. In Figure~\ref{f:ill_comp}, we show environment versus mass-weighted formation redshift for observed massive galaxies in our sample and for Illustris analogs (grey area). Given the small field of view of the simulation, and the consequent lack of very massive galaxies, the actual values of the number density in the simulation are very different from those obtained for the galaxies in our sample. For this reason, in the comparison we express the environment as the ratio between the surface mass density of simulated and observed galaxies and their corresponding mean values. Figure~\ref{f:ill_comp} seem to indicate the same trend with environment and mass-weighted redshift in observed data and in the Illustris simulation, even though Illustris galaxies do not reach the overdensity levels seen in the data. However, there are two effects to take into account. First, in the Illustris sample there are fewer galaxies than in the observed sample. The number of galaxies in the illustris simulation is $\sim$ 8 galaxies per bin below $\overline{z}_{M-w}$ 2 and $\sim$ 4 above $\overline{z}_{M-w}$ 2. Moreover, we have to remark that in the Illustris sample, two populations of galaxies are completely missing: galaxies with formation redshift $\overline{z}_{M-w} <$ 1.5 and galaxies with formation redshift $\overline{z}_{M-w} >$ 3.5. 
In our sample, galaxies formed at earlier times ($z_{M-w} \geq 2.5$) rarely live in environments denser than 2 times the average density, those regions are only inhabited by recently formed galaxies. Quantitatively, 40\% and 28\% of galaxies with $z_{M-w} < 1.5$ and $1.5 \leq z_{M-w} \leq 2.5$ are located in high density regions, twice the average number density.

Figure~\ref{f:ill_comp_z05} shows the same comparison as in Figure~\ref{f:ill_comp_z05}, but considering the redshift at which the galaxies formed 5\% of their mass, $z_{M-5\%}$. Illustris galaxies start forming stars at later times, the simulations do not have systems with $z_{M-5\%}>6$. Observed galaxies, on the other hand, start forming stars at earlier times. For observed galaxies, the trend between environment and formation redshift is still present when considering $z_{M-5\%}$ instead of $z_{M-w}$. This is not true for Illustris galaxies.

\subsection{Morphology of massive galaxies}\label{ss:morpho}

\begin{figure*}[ht]
\centering
\includegraphics[width=\linewidth]{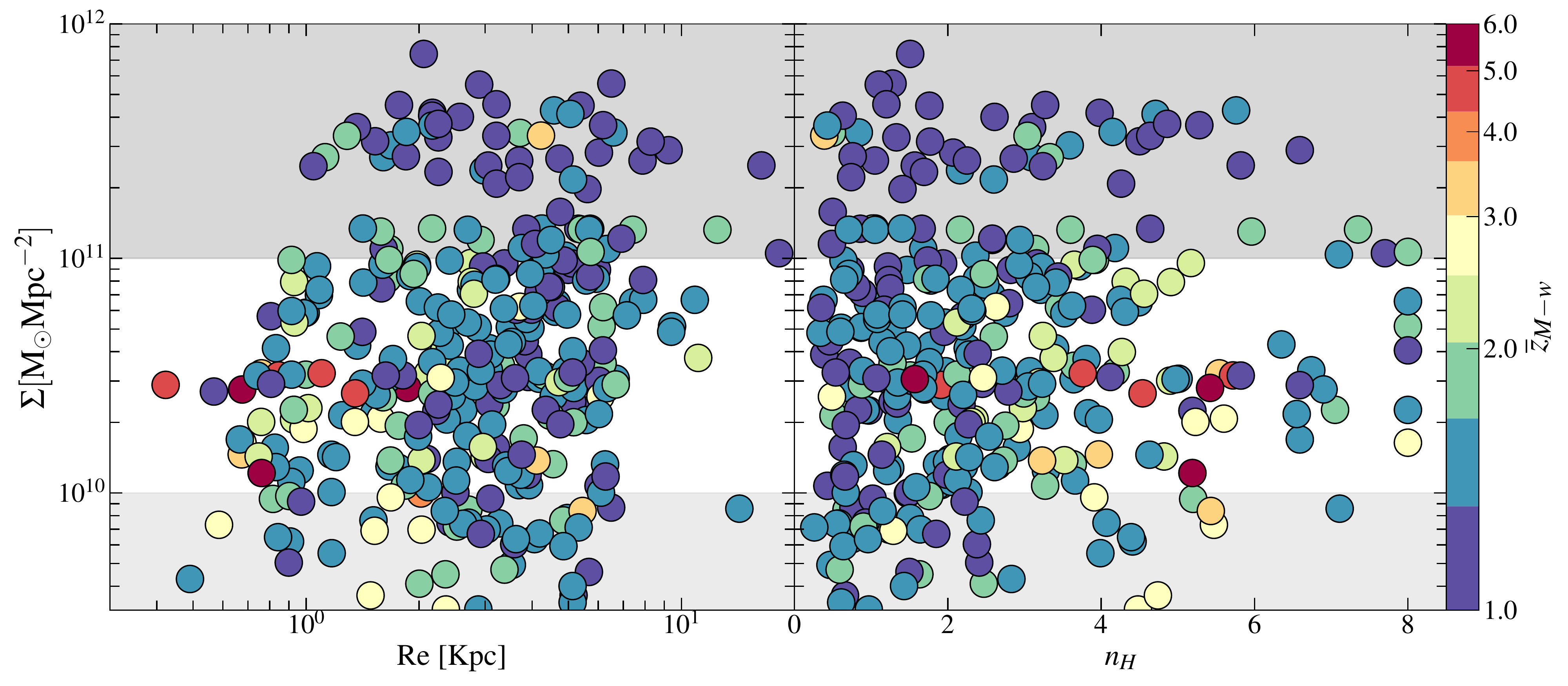} 
\caption{Surface mass density as a function of effective radius (left panel) and S\'ersic index (right panel) for galaxies in our sample. Points are color-coded according to their mass-weighted formation redshift.}
\label{f:env_size_n}
\end{figure*}

A detailed morphological analysis, and a 2D decomposition of the stellar population parameters of the galaxies in our sample could shed further light on the mechanisms intervening in their assembly. In \citet{vanderwel2012}, the parameters corresponding to a single Sersic fit \citep{sersic1968} performed with {\textit{galfit}} \citep{peng2010} were derived for all galaxies in our sample using the HST image in $\mathrm{H_{160}}$ band. 
In Figure~\ref{f:env_size_n}, we show the environment versus half-light radius and S\'ersic index.  Figure~\ref{f:env_size_n} shows that galaxies with very high mass-weighted formation redshifts are smaller with respect to more recently formed galaxies. 
Moreover, it shows that galaxies with higher mass-weighted formation redshift have higher S\'ersic index. This is quantified in Table~\ref{t:properties_zform}. Recently formed galaxies with $z_{M-w}<1.5$ have a median S\'ersic index of $n=1.8_{0.7}^{3.9}$, while galaxies with $z_{M-w} \geq 2.5$ have a median S\'ersic index of $n=3.8_{1.5}^{5.5}$. 
The trend with environment is much milder. Galaxies in the lowest density environment have a median S\'ersic index of $n=1.5_{0.7}^{4.1}$, while galaxies in the highest density environment have a median S\'ersic index of $n=2.2_{0.9}^{4.6}$. 
This is quantified in Table~\ref{t:env}, where we list the 16th, 50th, and 84th percentiles of the distribution of galaxies properties in the three environment.  

\begin{figure}[ht]
\hspace{-0.5cm}
 \includegraphics[width=0.5\textwidth]{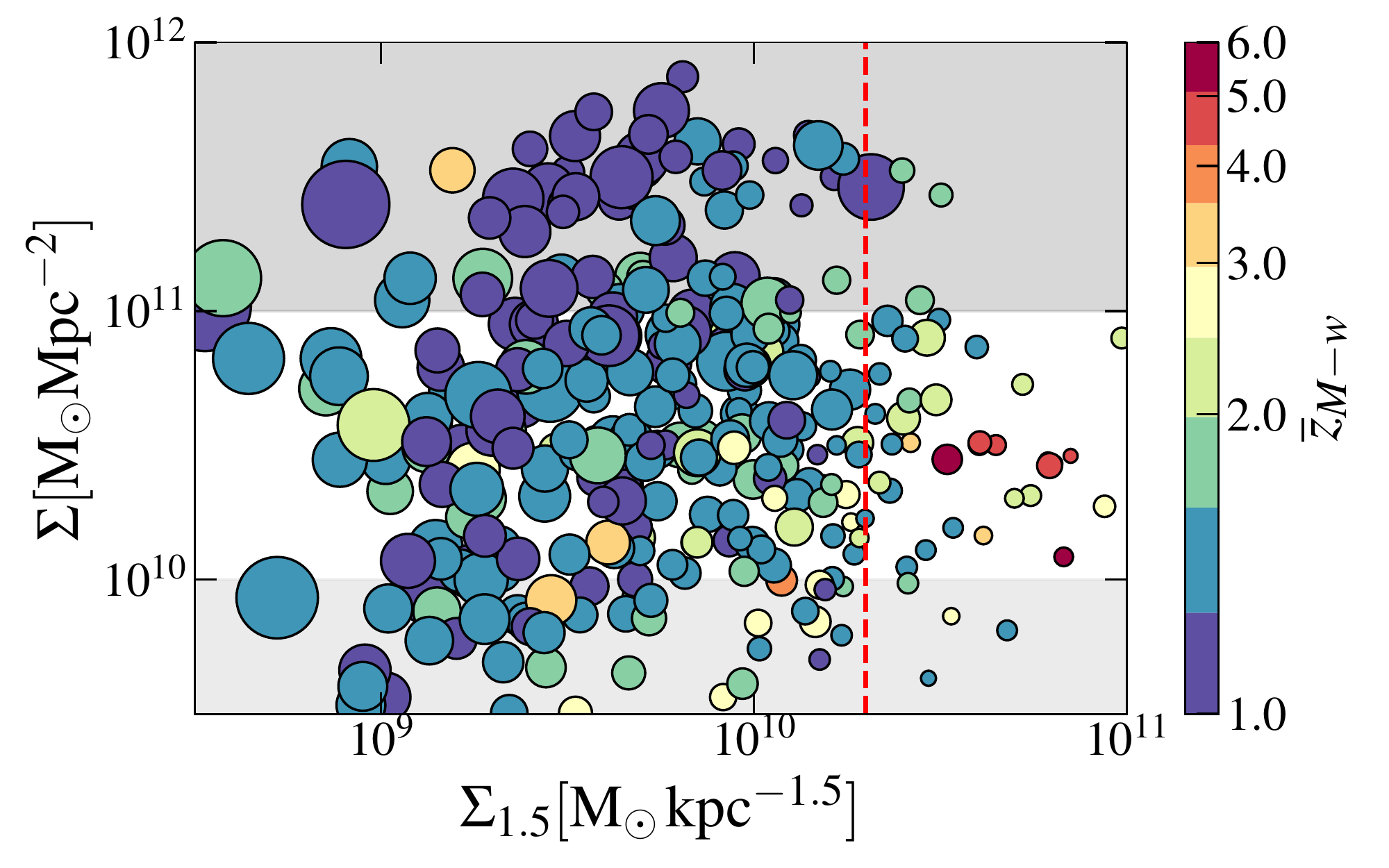} 
 \caption{Environment vs galaxy mass density for all galaxies in our sample. The lighter and darker grey areas represent the lowest and highest density environments identified by the 16th and 84th percentile of the distribution of the environment surface mass density. The size of the points is proportional to the effective radius of the galaxies. The points are color-coded according to the mass-weighted formation redshift. The red line marks the threshold in $\Sigma_{1.5}$ above which a galaxy is considered compact \citep{barro2013}.}
\label{f:env_den}
\end{figure}

\noindent Finally, in Figure ~\ref{f:env_den}, we show the environment versus their mass density, defined by its mass density ($\mathrm{\Sigma_{1.5}=MR_e^{-1.5}}$) following \citet{barro2013}, for our sample of massive galaxies. The dashed line in Figure ~\ref{f:env_den} represent the threshold at which galaxies are considered compact ($\mathrm{10^{10}\,M_{\odot}\,kpc^{-1.5}}$, \citealt{barro2013}). In total, 13\% of our sample can be identified with compact galaxies. This corresponds to a number density of: $\mathrm{3.0 \pm 0.4 \times 10^{-4} Mpc^{-3}}$. If we consider separately quiescent and star-forming compact galaxies, the densities are $\mathrm{2.5 \pm 0.4 \times 10^{-4} Mpc^{-3}}$ and $\mathrm{0.5 \pm 0.4 \times 10^{-4} Mpc^{-3}}$, respectively. These values are in agreement with the number densities of compact quiescent and star-forming galaxies at similar redshift \citep[$2.4 \pm 0.4$ $\mathrm{Mpc^{-3}}$, and $0.7 \pm 0.2$ $\mathrm{Mpc^{-3}}$, respectively, ][]{barro2013}. 

As we can see in Figure ~\ref{f:env_den}, compact galaxies are predominantly formed by old and mid-age galaxies (5\% and 57\%, respectively), while only 9\% of compact galaxies are young. Moreover, among old galaxies, 50\% of them are compact, while, in contrast this is true only for 17\% and 2\% of mid-age and young galaxies, respectively.
On the other hand, Figure ~\ref{f:env_den} shows that compact galaxies almost completely avoid the highest density environment (darker shaded area in the plot), contrarily to their low redshift analogs \citep[][]{baldry2021}. 
We checked that the same results are obtained using all three definitions of environment defined in Sect.\ref{ss:env_def}. 

\begin{deluxetable}{lcccc}
	\tablewidth{0pt}
	\tablecaption{50, 16th (lower) and 84th (upper) percentiles of the distribution of galaxy properties (same as in Table~\ref{t:properties}) for galaxies in different environments.
		\label{t:env}}
	\tablehead{\colhead{} & \colhead{Low density}  & \colhead{Intermediate densities} & \colhead{High density}}
	\startdata 
	Parameter & & &\\
	\hline
	Number & 52 & 225 & 55 \\
	Fraction & 15\% & 64\% & 21\%\\
	$z$ & $1.28_{1.10}^{1.42}$ & $1.22_{1.08}^{1.41}$ & $1.02_{1.01}^{1.24}$\\
	$\mathrm{\overline{t}_{M-w}\, [Gyr]}$ & $\mathrm{0.4_{0.1}^{1.8}}$ & $\mathrm{0.7_{0.2}^{1.6}}$ & $\mathrm{0.5_{0.2}^{1.4}}$\\
	$\mathrm{log(M/M_{\odot})}$ & $\mathrm{10.22^{10.41}_{10.08}}$ & $\mathrm{10.46_{10.15}^{10.75}}$ & $\mathrm{10.49_{10.17}^{10.80}}$ \\
	$\tau$ [Myr] & $\mathrm{252_{100}^{1473}}$ & $\mathrm{159_{100}^{631}}$ & $\mathrm{158_{100}^{501}}$ \\
    $\overline{z}_{M-w}$ & $\mathrm{1.47_{1.26}^{2.00}}$ & $\mathrm{1.47_{1.25}^{2.04}}$ & $\mathrm{1.23_{1.08}^{1.55}}$\\
    $\overline{z}_{M-5\%}$ & $\mathrm{1.52_{1.15}^{3.76}}$ & $\mathrm{1.52_{1.24}^{2.42}}$ & $\mathrm{1.29_{1.09}^{1.68}}$\\
    $\mathrm{SFR\, [M_{\odot} yr^{-1}]}$ & $\mathrm{21_{2}^{42}}$ & $\mathrm{18_{2}^{55}}$ & $\mathrm{23_{2}^{56}}$\\
	$\mathrm{A_V\, [mag]}$ & $\mathrm{1.3_{0.9}^{2.0}}$ & $\mathrm{1.4_{0.5}^{2.1}}$ & $\mathrm{1.2_{0.3}^{2.3}}$\\
	$\mathrm{Z\, [Z_{\odot}]}$ & $\mathrm{0.4_{0.2}^{2.5}}$ & $\mathrm{0.4_{0.2}^{2.5}}$ & $\mathrm{1.0_{0.2}^{2.5}}$\\
	$n\, \mathrm{[Mpc^{-2}]}$ & $0.32_{0.16}^{0.62}$ & $0.75_{0.40}^{1.60}$ & $3.42_{1.35}^{5.09}$\\
	$\mathrm{\Sigma\, [10^{10} M_{\odot}Mpc^{-2}]}$ & $0.7_{0.4}^{0.8}$ & $3.2_{1.4}^{7.1}$ & $24.9_{12.1}^{38.6}$\\
	$n_3\, \mathrm{[Mpc^{-2}]}$ & $0.04_{0.01}^{0.08}$ & $0.13_{0.04}^{0.57}$ & $1.85_{0.55}^{6.30}$\\
	S\'ersic index $n$ & $\mathrm{1.5_{0.7}^{4.1}}$ & $\mathrm{2.3_{0.9}^{4.3}}$ & $\mathrm{2.2_{0.9}^{4.6}}$\\
    Re [kpc] & $\mathrm{2.70_{1.17}^{5.13}}$ & $\mathrm{2.91_{1.16}^{5.14}}$ & $\mathrm{3.21_{1.71}^{5.78}}$\\
	\enddata
\end{deluxetable}

\section{Discussion} \label{s:disc}

Our dataset and analysis allow us to explore, for the first time at this redshift $1.0<z<1.5$,  the correlation between age and environment for galaxies above $10^{10}M_{\odot}$. We find that the majority ($\sim 80 \%$) of our sample consists in young ($\mathrm{\overline{t}_{M-w} < 0.5\,Gyr}$), star-forming ($\mathrm{\sim 24\,M_{\odot}/yr}$), jointly with recently formed galaxies (less than 2 Gyr prior to observation). A small ($<10\%$) fraction of galaxies with very old ages ($\overline{t}_{M-w} > 2$ Gyr) and low SFR ($\sim 3 M_{\odot} yr^{-1}$) was also identified. \\
We have identified at $1.0<z<1.5$ the very first massive galaxies ever formed ($z_{M-w}>2.5$), which arrived earlier to a quiescent state. These galaxies represent $\sim$ 10\% of our sample. 
According to our SFHs derived for these galaxies with spectrophotometric data covering the rest-frame optical, the first episodes of star formation in massive galaxies were able to assemble 5\% of their mass before $z\sim10$. The SFH of these first massive peaks at $z\sim10$, where they reached SFRs of about 30~$\mathrm{M}\,\mathrm{yr}^{-1}$. This first epoch of the formation of massive galaxies is able to assemble around 50\% of the total stellar mass (in place at $z\sim1$) by $z\sim5$.\\
Our results point out that those first-to-form massive galaxies do not live in the densest environments. By $z\sim1$ they clearly differentiate from younger, more active galaxies, which do reside in more populated regions (10 times more neighbors, 3 times more mass around) and are around 0.2~dex less massive. Our work directly probes the environment at $1.0<z<1.5$, but the results also indicate that the first episodes of star formation in massive galaxies could not happen in galactic environments as dense as those that are hosting most of the star formation activity at $z\sim1$, given that both the number and mass surface densities of neighbors are lower for the older galaxies at $1.0<z<1.5$. We also find that the stronger evolution happens below $z<1.3$, galaxies with different ages above this redshift present more similar environments, although follow the same trend of less dense environments for older galaxies.

We do not find significant differences in the efficiency of the most recent star formation as a function of environment, as revealed by very similar main sequences for galaxies in low and high density regions. However, we do find that galaxies in the highest density environments present more pronounced  SFR peaks in their SFHs, reaching value well above 10~$\mathrm{M_{\odot}}\,\mathrm{yr}^{-1}$. In contrast, the SFHs for galaxies in the lowest density environments peak at 1.5-2.0 lower SFRs and count with a more extended, longer timescale, slowly increasing evolution at earlier epochs. The environment would not affect, thus, the start of the star formation activity in the galaxies formed first, but would enhance the activity, for shorter timescales at $z<1.5$, coinciding with the peak of the cosmic SFR density, before the decline at $z<1$. 

Finally, there is yet another difference between the first massive galaxies to form, and those arriving later. The galaxies formed first tend to have more concentrated stellar mass internal distributions, they are distinctively more compact. This result is also consistent with our analysis of bulge properties in $z<1$ massive galaxies \citep{costantin2021,costantin2022}, so we are being able to probe the first wave of massive galaxy formation, which does not seem to be ruled by environment, and the second wave and disk formation era which extends to lower redshifts and suffer a more significant impact from the surrounding galaxy density.

To interpret this result, one has to consider different evolutionary paths that massive galaxies can follow. In particular:
\begin{itemize}
    \item galaxies formed later than $\mathrm{\overline{z}_{M-w}=2.5}
    $ are more likely to survive when they are accreted in proto-clusters or cluster-like environments \citep{bahe2019}. 
    \item Galaxies formed earlier ($\mathrm{\overline{z}_{M-w}>2.5}$) could:
    \begin{enumerate}
        \item evolve in isolation up to z=1.5; 
        \item evolve through a series of minor and even major merger events, cleaning out their close environment;
        \item be accreted to a protocluster-like system and then either get disrupted through gravitational or baryonic interactions with other galaxies, or get rejuvenated. 
    \end{enumerate}
\end{itemize}

Galaxies evolving through a series of minor \citep{walker1996} or major mergers \citep{toomre1972} will clear their immediate neighborhood, by either destroying the nearby galaxies with which they interact, stripping them below the detection limit of the sample, or ejecting them to long distances. Galaxies evolving in this way may be expected to have an early type morphology, and their environment would not be very dense because they cleared it, unless new galaxies continue falling to their vicinity.
\citet{behroozi2019}, using numerical simulations, found that in the stellar mass range between $M_{\star}=\mathrm{10^{10}\,- \, 10^{11}\, M_{\odot}}$ and at $z\sim1$, 10-20\% of galaxies have gone through a rejuvenation process. \citet{chauke2019} used spectroscopic data available for galaxies in the COSMOS field and found a similar fraction to that in \citet{behroozi2019} of rejuvenated galaxies among massive ($\mathrm{> 10^{10} M_{\odot}}$) systems at $\mathrm{z \sim 0.8}$.  If galaxies have indeed gone through a rejuvenation process, the SED techinque applied to their integrated photometry could  preferentially fit the $M_{∗}/L$ of the younger stellar population, missing older components \citep[][and references therein]{sorba2018}. This would also result in an underestimate of the mass of these galaxies, by a factor dependent on the galaxy’s specific star formation rate (sSFR). \\
According to \citet{bahe2019}, the survival of galaxies when they are accreted in a very dense region, like a cluster or proto-cluster environment, depend on the accretion redshift, $\mathrm{z_{acc}}$. According to their findings, overall, less than 50\% of galaxies with masses above $\sim 10^{10}M_{\sun}$ survive to the present day after they have been accreted into a massive halo. This fraction is dependent on galaxy mass, halo mass and strongly depends on the accretion redshift. Generally, at higher $z_{acc}$, the survival fraction drops steeply below 50\% and becomes negligible for $z_{acc} \gtrsim 4$. Below the scale of massive clusters, the drop into survival fraction occurs at lower $z_{acc}$ for higher-mass galaxies and in lower-mass hosts. For a fixed host mass, the lowest-mass galaxies are therefore the most likely to survive to $z=0$.
Our findings in terms of interplay between ages and environment could be summarized by hypothesizing that young galaxies with lower mass-weighted formation redshift are those that have probably been accreted more recently into the high density regions. Their star formation is enhanced due to the interaction with their many new neighbors and the availability of large amounts of gas. In contrast, 
older galaxies with higher mass-weighted formation redshift, are also probably accreted into massive structures earlier. As a result, they could be disrupted.
Old galaxies in our sample then, are those galaxies that either evolved in isolation, or are the central galaxies in their haloes and have already cleared our their environment to some extent.
However, as the oldest galaxies are the most compact in our sample, this could imply that this galaxies have indeed evolved in isolation.\\
These results apparently contradict previous works. For example, \cite{thomas2005} found that massive galaxy in high-density environments, i.e. cluster galaxies, are on average 2 Gyr older the low-density environment. \cite{webb2020} found that there is an age difference of $\sim$0.3 Gyr between galaxies in cluster and in the field for galaxies within $10^{10}-10^{11.8} M_{\sun}$. \cite{sobral2022} found that, at fixed stellar mass, quiescent galaxies in overdensity are older than quiescent galaxies in less dense environment. \cite{harshan2021} found that, using IllustrisTNG simulations, a small difference in the SFH of galaxies in proto-clusters and in the field can be found only for galaxies above $10^{10.5} M_{\sun}$, with cluster galaxies forming slightly more galaxies in the first 2 Gyr of the universe than field galaxies. In these studies these differences are mainly driven by higher mass galaxies (above $10^{10.5} M_{\sun}$) which are present in smaller number in this work. Moreover, it is important to point out that, in this work, we do not reach overdensities comparable to those present in clusters. 
\section{Conclusions}\label{s:concl}

In this paper, we have studied the interplay between environment and star-formation histories in massive ($\mathrm{log(M/M_{\odot})\geq 10}$) galaxies ay $1.0<z<1.5$ selected from the SHARDS spectro-photometric survey in GOODS-N. Thanks to the SHARDS data, and the complementary grism data from the CANDELS survey, we were able to characterize the SFHs of these galaxies. Moreover, thanks to the unprecedented accuracy on photometric redshift derived from the combination of SHARDS and grism data, we were able to characterize the environment in which these galaxies reside. Our results can be summarized as follows:
\begin{itemize}
    \item The majority of our sample (90\%) is formed by young (mass-weighted ages $\overline{t}_{M-w} <\ 0.5$ Gyr) and mid-age ($0.5\leq\overline{t}_{M-w} \leq 2.0$ Gyr) galaxies.
    \item There is a small fraction (10\%) of old galaxies ($\overline{t}_{M-w} \geq 2$ Gyr). 
    \item Young galaxies have lower masses, higher star formation rates, shorter formation time-scales and lower mass-weighted formation redshift ($\overline{z}_{M-w}$), while old galaxies have lower star formation rates, higher masses and longer formation time-scales and high mass-weighted formation redshift.
    \item In terms of environment, all massive galaxies avoid very low-density regions ($\mathrm{\Sigma < 10^{10}M_{\odot}Mpc^{-2}}$). 
    \item Very high density regions ($\mathrm{\Sigma = 10^{11} M_{\odot} Mpc^{-2} }$) are occupied by galaxies with lower mass-weighted formation redshift. Only 3\% of galaxies with $\overline{z}_{M-w} > 2.5$ can be found in the densest environments. The median surface mass density for galaxies with $\overline{z}_{M-w} > 2.5$ is:  $\Sigma = \mathrm{3.0_{0.8}^{3.2}\times 10^{10}M_{\odot}Mpc^{-2}}$.
    \item The average SFH of galaxies in lower density environment extends to higher redshift ($z=5$), while the average SFH of galaxies in the highest density environment does not extend beyond $z = 2$.
    \item We do not find significant differences in the efficiency of the most recent star formation as a function of environment. However, galaxies in the highest density environments present more pronounced  SFR peaks in their SFHs, reaching values above 50~$\mathrm{M_{\odot}}\,\mathrm{yr}^{-1}$. 
    \item There is a hint of a similar trend in the Illustris simulation. However, the distribution of mass-weighted redshift of Illustris galaxies is much narrower than the observed one. Galaxies with low ($\overline{z}_{M-w}<$1.5) and high ($\overline{z}_{M-w}>$3.5) formation redshifts are missing from the simulation. 
    \item Galaxies formed first tend to have more concentrated stellar mass internal distributions, they are distinctively more compact.
\end{itemize}

\section*{Acknowledgments}
MA acknowledges financial support from Comunidad de Madrid under Atracci\'on de Talento grant 2020-T2/TIC-19971. MA and PGPG acknowledge support from Spanish Ministerio de Ciencia, Innovaci\'on y Universidades through grant PGC2018-093499-B-I00. A. Garc\'ia-Argum\'anez acknowledges the support of the Universidad Complutense de Madrid through the predoctoral grant CT17/17-CT18/17
This work has made use of the Rainbow Cosmological Surveys Database, which is operated by the Centro de Astrobiolog\'ia (CAB/INTA), partnered with the University of California Observatories at Santa Cruz (UCO/Lick,UCSC).

\section*{Data Availability}
The data underlying this article will be shared on reasonable request to the corresponding author.
\bibliography{main}
\bibliographystyle{aa}

\appendix
\section{SED fitting} \label{a:sed}
\begin{figure*}[h]
  \hspace*{-0.35cm}%
\includegraphics[scale=0.55]{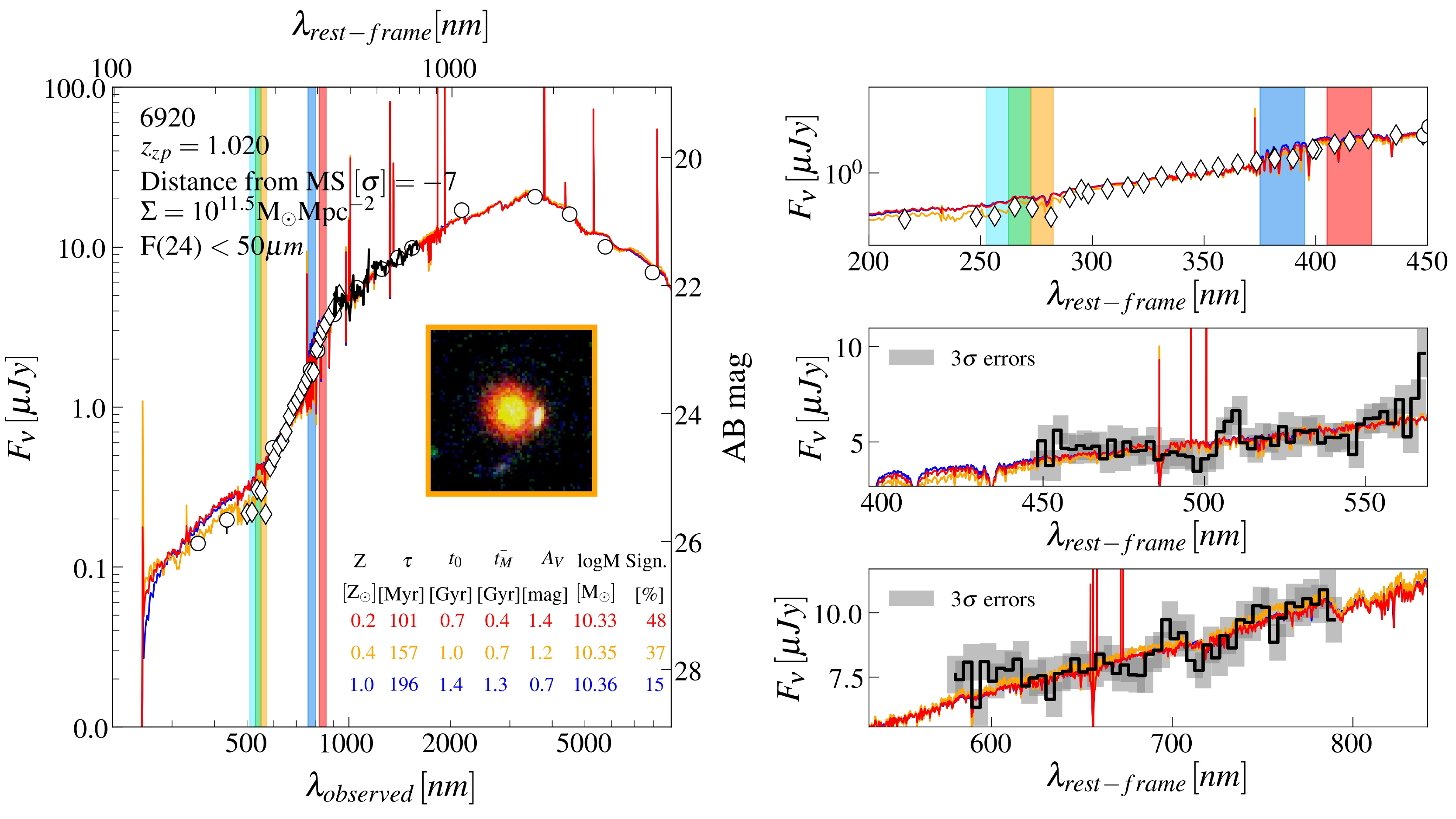}
\caption{Example of our SED-fitting method. Left panel: SED fitting of one of our galaxies. For this galaxy, three clusters of solutions were obtained (red, orange, and blue lines). The white circles represent the SHARDS spectrophotometric data and the diamonds are broad-band data. The G102 and G141 spectra are plotted as black lines, while the shaded grey areas represent the 3$\sigma$ errors. The coloured areas represent the bands used to determine the MgUV and D4000 indices. The best-fitting parameters of the three models are shown in the legend, including metallicity, star formation time-scale, age, mass-weighted age, dust attenuation, stellar mass, and statistical significance of the solution. Top right panel: zoom in the SHARDS region. The errors are smaller than the symbols. Right central panel and right bottom panel: zoom in the regions corresponding to G102 and G141, respectively. }
\label{f:sed_example}
\end{figure*}

\begin{figure}
\hspace{-1cm}
\begin{tabular}{@{}cc@{}}
\hspace{-0.5cm}
 \raisebox{-\height}{\includegraphics[width=10.cm,trim= 0 5 5 -20,clip]{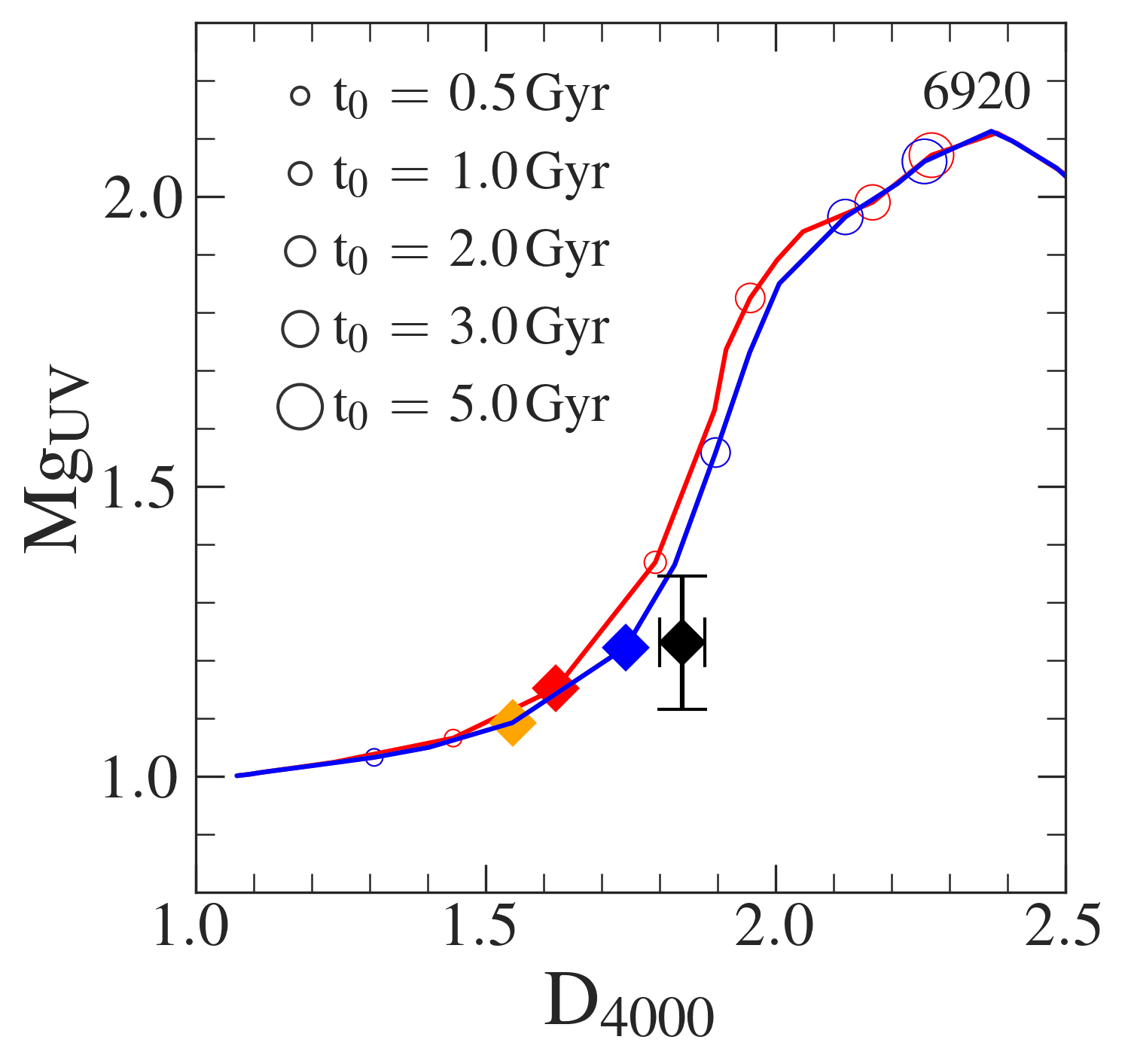}}
    \begin{tabular}[t]{@{}cc@{}}
    \hspace{-1.cm}
    \raisebox{-\height}{\includegraphics[width=0.25\textwidth, trim= 0 5 5 -5,clip]{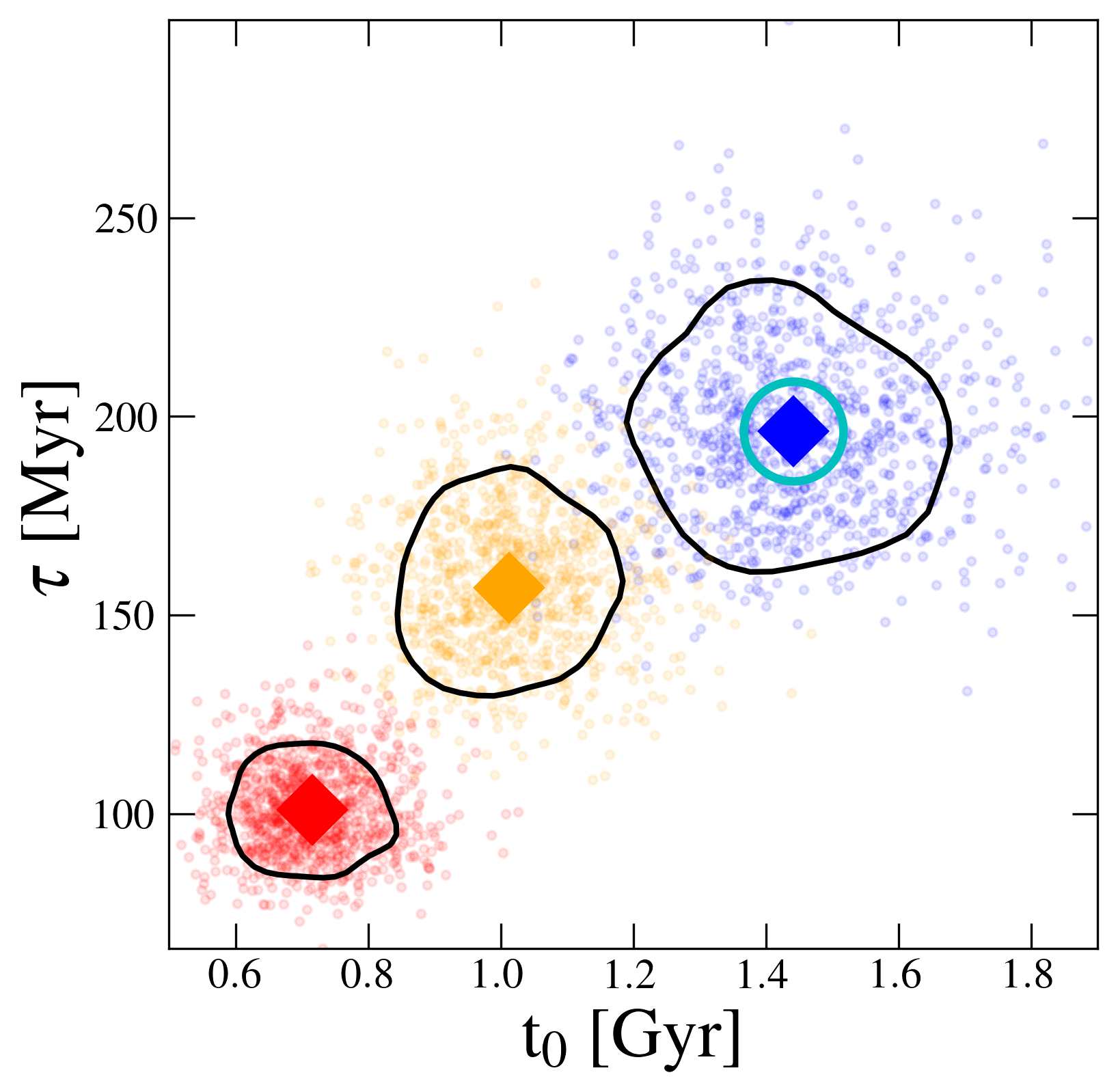}} & 
        \hspace{-0.3cm}
        \raisebox{-\height}{\includegraphics[width=0.25\textwidth, trim= 0 5 5 -5,clip]{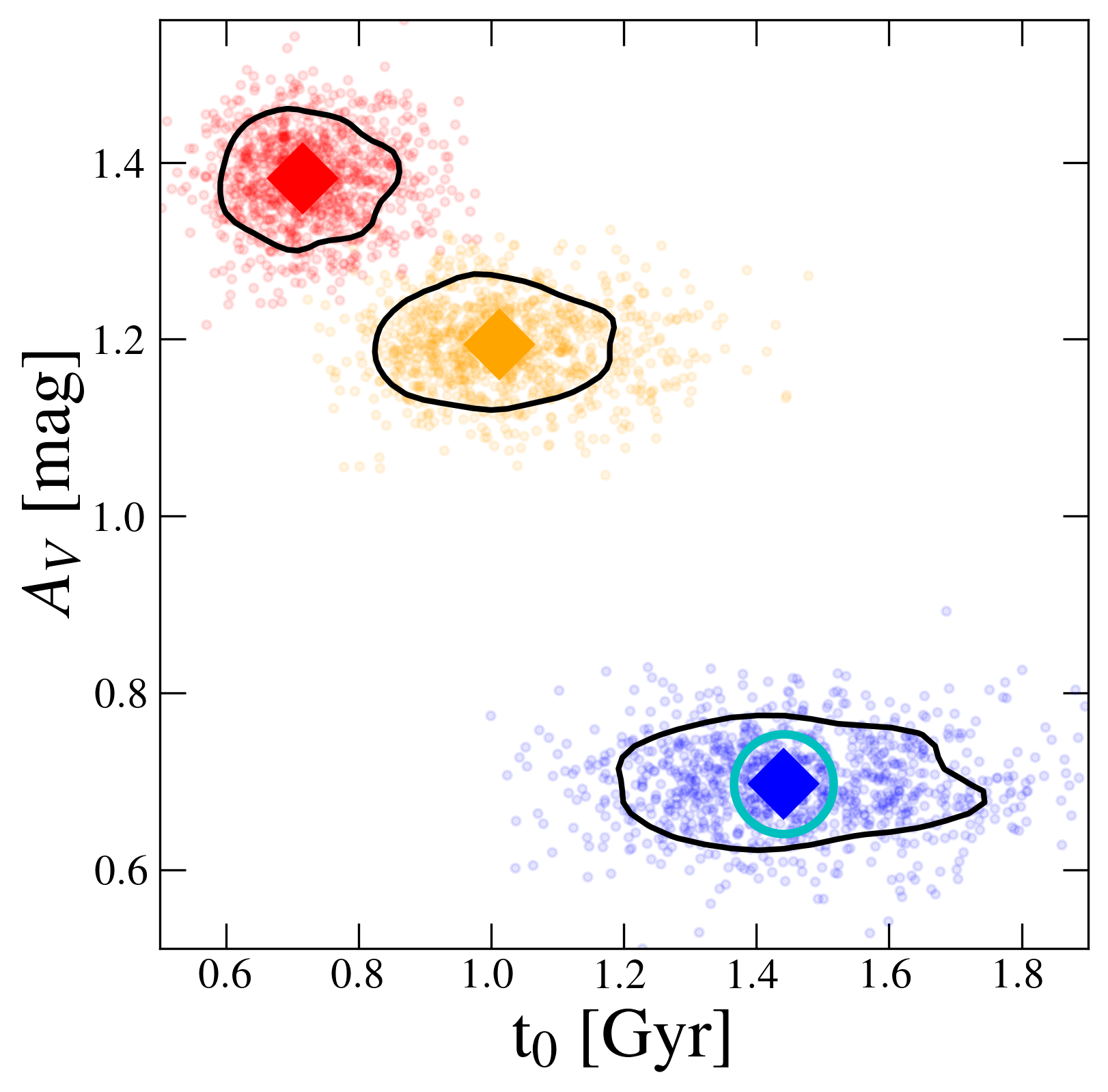}} \\[1.8cm]
        (b) & (c) \\
            \hspace{-1.cm}
        \raisebox{-\height}{\includegraphics[width=0.26\textwidth, trim= 0 5 5 -5,clip]{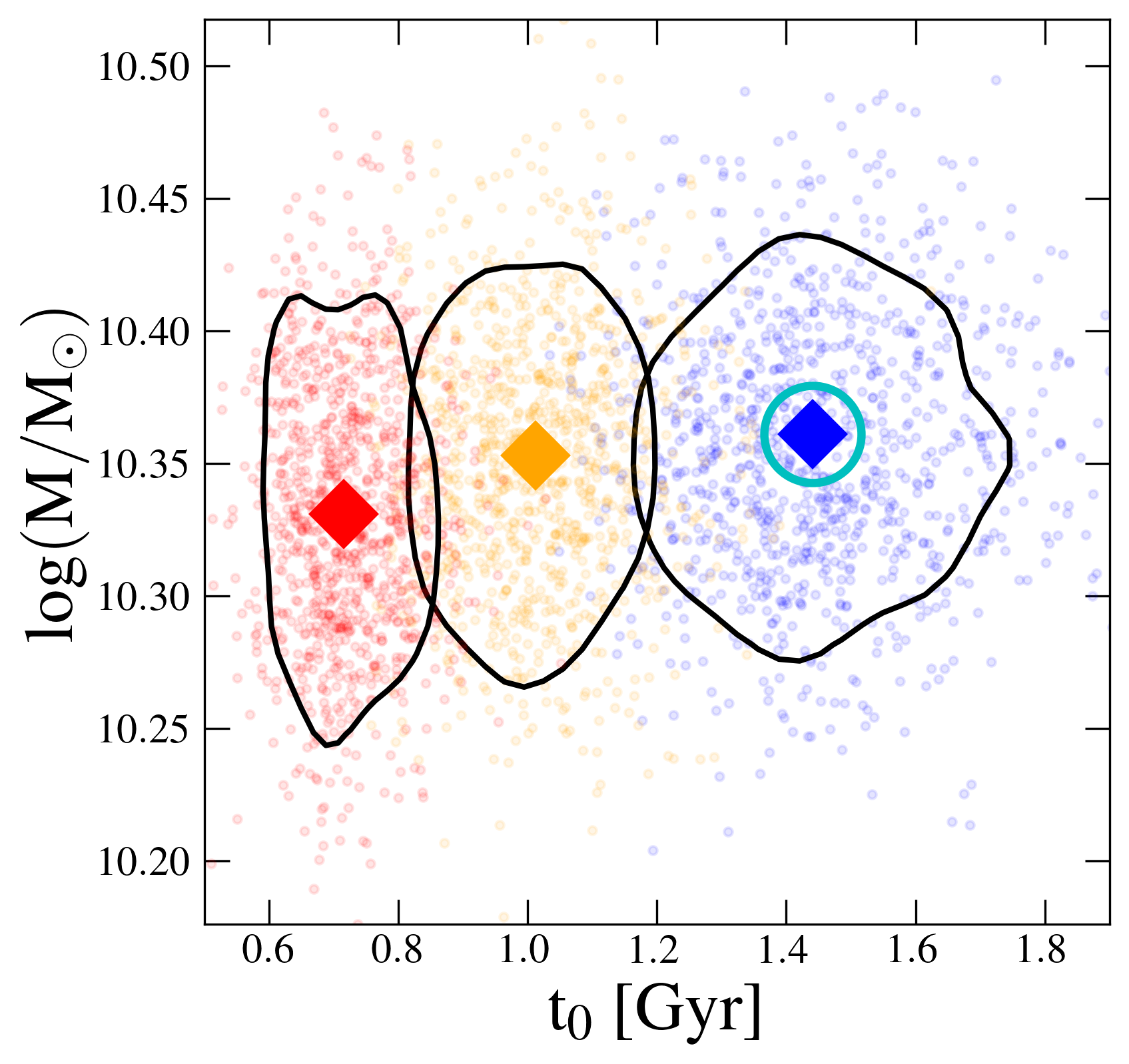}} & 
            \hspace{-0.3cm}
        \raisebox{-\height}{\includegraphics[width=0.25\textwidth, trim= 0 5 5 -5,clip]{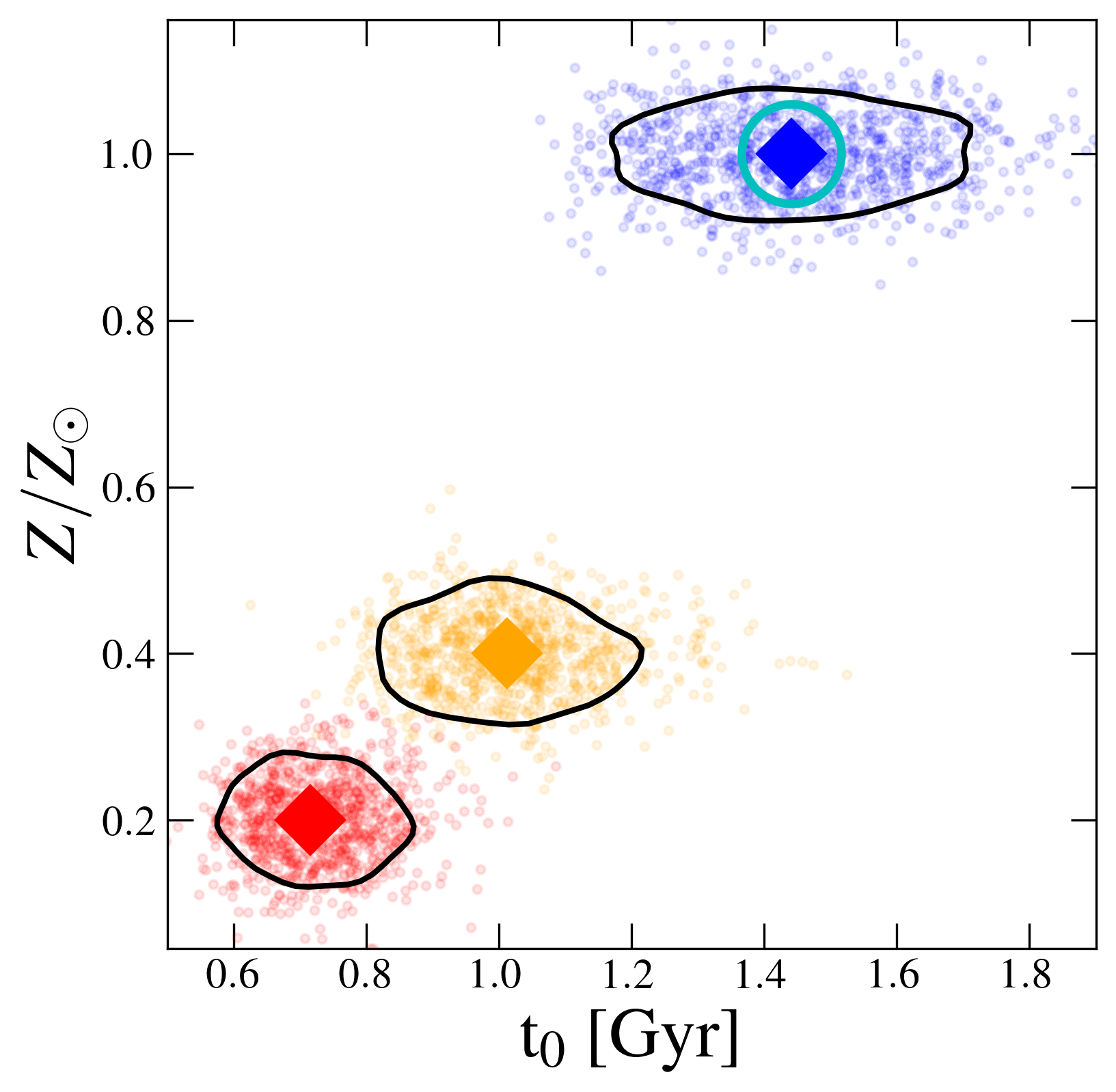}} \\
      (d) & (e) 
    \end{tabular}\\
\hspace{-8.5cm}  (a)\\
\end{tabular}
    \caption{Panel (a): evolutionary tracks in the MgUV–D4000 plane for the three best-fitting model. The open circles show the expected values of the indices at different ages, $\mathrm{t_0}$ (from smaller to larger symbols: 0.5, 1.0, 2.0, 3.0, and 5.0 Gyr). The large coloured diamonds show the location of the indices at the age of the best-fitting solution. The black symbol represents the indices measured from the SHARDS photometry. Panels (b - e): Distribution of the 1000 best-fitting solutions in the age/time-scale (Panel a), age/extinction (Panel b), age/mass (Panel c) and age/metallicity (Panel d) planes. The large diamond represent the clusters mean solutions. Black contours enclose 68\% of the solutions in each cluster. In cyan the solution adopted for this galaxy. }
    \label{f:tracks}
\end{figure}

\noindent In this section, we show an example of our SED fitting method. In Fig.~\ref{f:sed_example}, we depict the SED and fitting models for one of the galaxies in our sample. The fitting templates are matched to the spectral resolution of the grism data. For this galaxy, we find three clusters of solutions, shown in red, orange and blue from the most significant to less significant ones, shown in Fig.~\ref{f:tracks}(a). 

In Fig.~\ref{f:tracks}(a), we show the $Mg_{UV}$ and $D4000$ indices for the same galaxy shown in Fig.~\ref{f:sed_example}, together with the  tracks expected for the evolution of these indices according to the models of the two best-fitting (degenerate) solutions obtained for that galaxy. We note that the 3 different solutions found are linked to a $\tau$-$t_0$ degeneracy, as shown in Fig.~\ref{f:tracks}(b), linked to other degeneracies involving the metallicity and the attenuation. The most statistically significant solution (in red, with a significance of 48\%) is not compatible with the value of the two indices, hence we discard it. Moreover, this galaxy is not detected in the IR. The first two solutions, with attenuations larger than $\mathrm{A_V}>1$~mag, predict a 24~$\mu$m flux of $\sim$ 95 and 73 $\mu$Jy. Therefore,
we rejected the first two solutions and chose as the best solution the third one, which predicts a 24~$\mu$m flux of 36~$\mu$Jy, below the detection limit of our MIPS data, and presents the closest index values to those measured with SHARDS data.

\end{document}